\documentclass[useAMS,usenatbib]{mn2e}
\usepackage{soul}
\usepackage{amssymb}
\usepackage{amsmath}
\usepackage{float}
\usepackage{times}
\usepackage{graphicx}
\usepackage{multirow}
\usepackage{tikz}

\title[Clustered Radio Interferometric Calibration]{Clustered Calibration: An Improvement to Radio Interferometric Direction Dependent Self-Calibration}

\author[Kazemi et al. ]{ S. Kazemi$^{1}$\thanks{E-mail:
kazemi@astro.rug.nl}, S. Yatawatta$^{2}$, S. Zaroubi$^{1}$ \\ 
$^{1}$Kapteyn Astronomical Institute, University
of Groningen, P.O. Box 800, 9700 AV Groningen, the Netherlands\\
$^{2}$ASTRON, Postbus 2, 7990 AA Dwingeloo, the Netherlands }

\begin{document}

\pagerange{\pageref{firstpage}--\pageref{lastpage}} \pubyear{2007}

\maketitle 

\label{firstpage}

\begin{abstract}
The new generation of radio synthesis arrays, such as LOFAR and SKA, have been designed to surpass existing arrays in terms of sensitivity, angular resolution and frequency coverage. This evolution has led to the development of advanced calibration techniques that ensure the delivery of accurate results at the lowest possible computational cost. However, the performance of such calibration techniques is still limited by the compact, bright sources in the sky, used as calibrators. It is important to have a bright enough source that is well distinguished from the background noise level in order to achieve satisfactory results in calibration. This paper presents ``clustered calibration'' as a modification to traditional radio interferometric calibration, in order to accommodate faint sources that are almost below the background noise level into the calibration process. The main idea is to employ the information of the bright sources' measured signals as an aid to calibrate fainter sources that are nearby the bright sources. In the case where we do not have bright enough sources, a source cluster could act as a bright source that can be distinguished from background noise. For this purpose, we construct a number of source clusters assuming that the signals of the sources belonging to a single cluster are corrupted by almost the same errors. Under this assumption, each cluster is calibrated as a single source, using the combined coherencies of its sources simultaneously. This upgrades the power of an individual faint source by the effective power of its cluster. The solutions thus obtained for every cluster are assigned to each individual source in the cluster. We give performance analysis of clustered calibration to show the superiority of this approach compared to the traditional un-clustered calibration. We also provide analytical criteria to choose the optimum number of clusters for a given observation in an efficient manner.
 \end{abstract}


\section{Introduction}\label{sec:introduction}
Low frequency radio astronomy is undergoing a revolution as a new generation of radio interferometers with increased sensitivity and resolution, such as the LOw Frequency ARray (LOFAR)\footnote{http://www.lofar.org},  the Murchison Wide-field Array (MWA) \footnote{http://www.mwatelescope.org} and the Square Kilometer Array (SKA)\footnote{http://www.skatelescope.org} are being devised and some are already becoming operational. These arrays form a large effective aperture by the combination of a large number of antennas using aperture synthesis \citep{A.R.1}. In order to achieve the full potential of such an interferometer, it is essential that the observed data is properly calibrated before any imaging is done. Calibration of radio interferometers refers to the estimation and reduction of errors introduced by the atmosphere and also by the receiver hardware, before imaging. We also consider the removal of bright sources from the observed data part of calibration, that enable imaging the weak background sources. For low frequency observations with a wide field of view, calibration needs to be done along multiple directions in the sky.  Proper calibration across the full field of view  is required  to achieve the interferometer's desired precision and sensitivity giving us high dynamic range images. 

Early radio astronomy used external (classical) calibration which estimates the instrumental unknown parameters using a radio source with known properties. This method has a relatively low computational cost and a fast execution time, but it generates results strongly affected by the accuracy with which the source properties are known. In addition, existence of an isolated bright source, which is the most important requirement of external calibration, in a very wide field of view is almost impractical, and even when it does, external calibration would only give information along the direction of the calibrator. The external calibration was then improved by self-calibration \citep{selfcal}. Self-calibration has the advantage of estimating both the source and instrumental unknowns, without the need of having a prior knowledge of the sky, only utilizing the instrument's observed data. Moreover, it can achieve a superior precision by iterating between the sky and the instrumental parameters.

The accuracy of any calibration scheme, regardless of the used technique, is determined by the level of Signal to Noise Ratio (SNR). This limits the feasibility of any calibration scheme to only bright sources that have a high enough SNR to be distinguished from the background noise \citep{Bernardi, Liu, Pindor}. Note that interferometric images are produced using the data observed during the total observation (integration) time. However, calibration is done at shorter time intervals of that total duration. This increases the noise level of the data for which calibration is executed compared to the one in the images. In other words, there are some faint sources that appear well above the noise in the images while they are well below  the noise level during calibration. It has been a great challenge to calibrate such faint sources having a very low SNR. In this paper we present clustered self-calibration, introduced in \citet{S.K.2}, and emphasize its better performance compared to un-clustered calibration below the calibration noise level. Existing un-clustered calibration approaches \citep{Intema, Oleg} can only handle a handful of directions where strong enough sources are present and we improve this, in particular for subtraction of thousands of sources over hundreds of directions in the sky, as in the case for LOFAR \citep{Bregman}.

The implementation of clustered calibration is performed by clustering the sources in the sky, assuming that all the sources in a single cluster have the same corruptions, and calibrating each cluster as a single source. At the end, the obtained calibration solutions for every source cluster is assigned to all the sources in that cluster. This procedure improves the information used by calibration by incorporating the total of signals observed at each cluster instead of each individual source's signal. Therefore, when SNR is very low, it provides a considerably better result compared to un-clustered calibration. Moreover, calibrating for a set of source clusters instead of for all the individual sources in the sky  reduces the number of directions along which calibration has to be performed, thus considerably cutting down the computational cost. However, there is one drawback in clustered calibration: The corruptions of each individual source in one given cluster will almost surely be slightly different from the corruptions estimated for the whole cluster by calibration. We call this additional error as ``clustering error'' and in order to get the best performance in clustered calibration, we should find the right balance between the improvement in SNR as opposed to degradation by clustering error.

Recently, clustering methods have gained a lot of popularity in dealing with large data sets \citep{Clustering}. There are various clustering approaches that could be applied to calibration based on their specific characteristics. An overview of different clustering methods is given in \citet{Clusteringbook}. Clustering of radio sources should take into account  (i) their physical distance from each other and (ii) their individual intensity. The smaller the angular separations of sources are, the higher the likelihood that they share the same corruptions in their radiated signals. Moreover, in order to get the best accuracy in the calibration results, there should be a balance between effective intensities of different clusters. Thus, in the clustering procedure, every source should be weighted suitably according to its brightness intensity.

The brightness distribution of radio source in the sky is a power law and the spatial distribution is Poisson. Therefore, clustering the sources via probabilistic clustering approaches is computationally complex. Hierarchical clustering \citep{Hierarchicalclustering} is a well-known clustering approach with a straight forward implementation suitable for our case. But, its computational cost grows exponentially with the size of its target data set which can be a disadvantage when the number of sources is huge. Weighted K-means clustering \citep{weightedk-means, kmeans} is also one of the most used of clustering schemes applicable in clustered calibration. The advantage of this clustering technique is its low computational cost, which is proportional to the number of clusters and the size of the target data set.  Nonetheless, we emphasize that the computational time taken by any of the aforementioned clustering algorithms is negligible compared with the computational time taken by the actual calibration. Therefore, we pursue all clustering approaches in this paper. However, the use of Fuzzy C-means clustering \citep{Fuzzyclustering} in clustered calibration requires major changes in the calibration data model and will be explored in future work.

This paper is organized as follows: First, in section \ref{Data Model} we present the general data model used in clustered calibration. In section \ref{clustering}, we present modified weighted K-means and divisive hierarchical clustering for clustering sources in the sky. Next, in section \ref{Performance Analysis}, we focus on analyzing the performance of clustered calibration, with an a priori clustered sky model, and compare the results with un-clustered calibration. In clustered calibration, there is contention between the improvement of SNR by clustering sources and the errors introduced by the clustering of sources itself. Thus, we relate the clustered calibration's performance to the effective Signal to Interference plus Noise Ratio (SINR) obtained for each cluster. For this purpose, we use statistical estimation theory and the Cramer-Rao Lower Bounds (CRLB) \citep{CRVB}. In section \ref{Model Order Selection}, we derive criteria for finding the optimum number of clusters for a given sky. We use the SINR analysis and adopt Akaike's Information Criterion (AIC) \citep{H.1} and the Likelihood Ratio Test (LRT) \citep{L.R.T} to estimate the optimum number of clusters. We present simulation results in section \ref{example} to show the superiority of clustered calibration to un-clustered calibration and the  performance of the presented criteria in detecting the optimum number of clusters. Finally, we draw our conclusions in section \ref{summary}. Through this paper, calibration is executed by the Space Alternating Generalized Expectation maximization (SAGE) \citep{J.A.1, S.2, S.K} algorithm. Moreover, in our simulations, radio sources are considered to be uniformly distributed in the sky and their flux intensities follow Raleigh distribution, which is the worst case scenario. In real sky models, there usually exist only a few (two or three) number of bright sources which dominate the emission. In the presence of these sources and the background noise, it is impractical to solve for the other faint sources in the field of view individually. Therefore, obtaining a better calibration performance via the clustered calibration approach, compared to the un-clustered one, is guaranteed. Therefore, in section \ref{example} we illustrate this using simulations in which the brightness distribution of sources is a power law with a very steep slope. On top of that, \citet{S.K.3, Y.NCP}  also present the performance of clustered calibration on real observations using LOFAR.

The following notations are used in this paper: Bold, lowercase letters refer to column vectors, e.g., {\bf y}. Upper case bold letters refer to matrices, e.g., {\bf C}.  All parameters are complex numbers, unless stated otherwise. The inverse, transpose, Hermitian transpose, and conjugation of a matrix are presented by $(.)^{-1}$, $(.)^T$, $(.)^H$, and $(.)^*$, respectively. The statistical expectation operator is referred to as $E\{.\}$. The matrix Kronecker product and the proper (strict) subset are denoted by $\otimes$ and $\subsetneq$, respectively. ${\bf I}_n$ is the $n\times n$ identity matrix and $\varnothing$ is the empty set. The Kronecker delta function is presented by $\delta_{ij}$. $\mathbb{R}$ and $\mathbb{C}$ are the sets of Real and Complex numbers, respectively. The Frobenius norm is shown by $||.||$. Estimated parameters are denoted by a hat, $\widehat{(.)}$. All logarithmic calculations are to the base $e$.  The multivariate Gaussian and Complex Gaussian distributions are denoted by $\mathcal{N}$ and $\mathcal{CN}$, respectively.

\section{Clustered Self-calibration Data Model}\label{Data Model}
In this section, we present the measurement equation of a polarimetric clustered calibration in detail \citep{J.P.1, J.P.2}. Suppose we have a radio interferometer consisting of $N$ polarimetric antennas where each antenna is composed of two orthogonal dual-polarization feeds that observe $K$ compact sources in the sky. Every $i$-th source signal, $i\in\{1,2,\ldots,K\}$, causes an induced voltage of $\widetilde{{{\bf{v}}}}_{pi}=[v_{Xpi}\ v_{Ypi}]^T$ at $X$ and $Y$ dipoles of every $p$-th antenna, $p\in\{1,2,\ldots,N\}$. In practice,
\begin{equation}
\widetilde{{{\bf{v}}}}_{pi}={{\bf{J}}}_{pi}{{\bf{e}}}_{i},\label{s1}
\end{equation}
where ${\bf{e}}_i=[e_{Xi}\ e_{Yi}]^T$ is the source's electric field vector and ${{\bf{J}}}_{pi}$ represents the $2\times 2$ Jones matrix \citep{J.P.1} corresponding to the direction-dependent gain corruptions in the radiated signal. These corruptions are originated from the instrumental (the beam shape, system frequency response, etc.) and the propagation (tropospheric and ionospheric distortions, etc.) properties which later on, in this section, will be explained in more detail.

The signal ${\bf v}_p$ obtained at every antenna $p$ is a linear superposition of the $K$ sources corrupted signals, $\widetilde{{{\bf{v}}}}_{pi}$ where $i\in\{1,2,\ldots,K\}$, plus the antenna's thermal noise. The multitude of ignored fainter sources also contributes to this additive noise. 

The voltages collected at the instrument antennas get corrected for geometric delays, based on the location of their antennas, and some instrumental effects, like the antenna clock phases and electronic gains. Then, they are correlated in the array's correlator to generate visibilities \citep{J.P.1}. The visibility matrix of the baseline $p-q$, $\operatorname{E}\{{\bf v}_p\otimes{\bf v}_q^H\}$, is given by
\begin{equation}
{\bf V}_{pq}={\bf G}_p \left(\sum_{i=1}^K {{\bf{J}}}_{pi}({\pmb{\theta}}){{\bf{C}}}_{i\{pq\}}{{\bf{J}}}^H_{qi}({\pmb{\theta}}) \right) {\bf G}_q^H +{\bf{N}}_{pq}.\label{s2}
\end{equation}
In Eq. (\ref{s2}), ${\pmb{\theta}}\in \mathbb{C}^P$, $P=4KN$, is the unknown instrumental and sky parameter vector, ${\bf N}_{pq}$ is the additive $2\times 2$ noise matrix of the baseline $p-q$, and ${{\bf{C}}}_{i\{pq\}}$ is the Fourier
transform of the $i$-th source coherency matrix ${{\bf{C}}}_{i}=\operatorname{E}\{{{\bf{e}}}_{i}\otimes{{\bf{e}}}_{i}^H\}$ \citep{bornwolf, J.P.1}. If the $i$-th source radiation intensity is $I_i$, then ${{\bf{C}}}_{i}=\frac{I_i}{2}{\bf I}_2$. Considering this source to have equatorial coordinate, (Right Ascension $\alpha$, Declination $\delta$), equal to $(\alpha_i,\delta_i)$, and the geometric components of baseline $p-q$ to be $(u,v,w)$, then 
\begin{equation}
{{\bf{C}}}_{i\{pq\}}=e^{-2\pi \jmath (ul+vm+w(\sqrt{1-l^2-m^2}-1))}{{\bf{C}}}_{i},\label{sa22}
\end{equation}
where $\jmath^2=-1$ and,
\begin{eqnarray*}
&&l=sin(\alpha_i-\alpha_0)cos(\delta_i),\\
&&m=cos(\delta_0)sin(\delta_i)-cos(\alpha_i-\alpha_0)cos(\delta_i)sin(\delta_0),
\end{eqnarray*}
are the source direction components corresponding to the observation phase reference of $(\alpha_0,\delta_0)$ \citep{A.R.1}. The errors common to all directions, such as the receiver delay and amplitude errors, are given by ${\bf G}_p$ and ${\bf G}_q$. Initial calibration at a finer time and frequency resolution is performed to estimate and correct  for ${\bf G}_p$-s and the corrected visibilities are obtained as
\begin{equation}
\widetilde{\bf V}_{pq}={\bf G}_p^{-1} {\bf V}_{pq} {\bf G}_q^{-H}.\label{ss3}
\end{equation}
The remaining errors are unique to a given direction, but residual errors in ${\bf G}_p$-s are also absorbed into these errors, which are denoted by ${\bf{J}}_{pi}$ in the usual notation.

Vectorizing Eq. (\ref{ss3}), the final visibility vector of the baseline $p-q$ is given by 
\begin{equation}
{\bf v}_{pq}=\sum_{i=1}^K {\bf J}^*_{qi}({\pmb{\theta}})\otimes{\bf J}_{pi}({\pmb{\theta}})\mbox{vec}({\bf C}_{i\{pq\}})+{\bf n}_{pq}.\label{s3}
\end{equation}
Stacking up all the cross correlations (measured visibilities) and noise vectors in ${\bf y}$ and ${\bf n}$, respectively, the un-clustered self-calibration measurement equation is given by
\begin{equation}
{\bf y}=\sum_{i=1}^K {\bf s}_i({\pmb{\theta}})+{\bf n}.\label{s4}
\end{equation}
In Eq. (\ref{s4}), ${\bf y},{\bf n}\in \mathbb{C}^M$, $M=2N(N-1)$, the noise vector is considered to have a zero mean Gaussian distribution with covariance ${\pmb{\Pi}}$, ${\bf n}\sim \mathcal{N}(0,{\pmb{\Pi}}_{M\times M})$, and the nonlinear function ${\bf s}_i(\pmb{\theta})$ is defined as
\begin{equation}
{\bf s}_i(\pmb{\theta})\equiv\left[\begin{array}{c}
{\bf J}^*_{2i}(\pmb{\theta})\otimes{\bf J}_{1i}(\pmb{\theta})\mbox{vec}({\bf C}_{i\{pq\}})\\
{\bf J}^*_{3i}(\pmb{\theta})\otimes{\bf J}_{1i}(\pmb{\theta})\mbox{vec}({\bf C}_{i\{pq\}})\\
\vdots\\
{\bf J}^*_{Ni}(\pmb{\theta})\otimes{\bf J}_{(N-1)i}(\pmb{\theta})\mbox{vec}({\bf C}_{i\{pq\}})\end{array}\right].\label{sa5}
\end{equation}

Calibration is essentially the Maximum Likelihood (ML) estimation of the unknown parameters  ${\pmb{\theta}}$ ($P$ complex values or $2P$ real values), or of the Jones matrices ${\bf{J}}({\pmb{\theta}})$, from Eq. (\ref{s4}) and removal of the $K$ sources. Note that calibration methods could also be applied to the uncorrected visibilities of (\ref{ss3}) to estimate ${\bf G}_p$ and ${\bf G}_q$ errors as well. 

The Jones matrix ${\bf{J}}_{pi}$, for every $i$-th direction and at every $p$-th antenna, is given as 
\begin{equation}
{\bf{J}}_{pi}\equiv {\bf{E}}_{pi}{\bf{Z}}_{pi}{\bf{F}}_{pi}.\label{i3}
\end{equation} 
In Eq. (\ref{i3}), ${\bf{E}}_{pi}$, ${\bf{Z}}_{pi}$, and ${\bf{F}}_{pi}$ are the antenna's voltage pattern, ionospheric phase fluctuation, and Faraday Rotation Jones matrices, respectively. In practice, the ${\bf{E}}$, ${\bf{Z}}$, and ${\bf{F}}$ Jones matrices obtained for nearby directions and for a given antenna are almost the same. Thus, for every antenna $p$, if the $i$-th and $j$-th sources have a small angular separation from each other, we may consider
\begin{equation}
{\bf J}_{pi}\cong{\bf J}_{pj}.\label{i4}
\end{equation}  
This is the underlying assumption for clustered calibration. 

Clustered calibration first assigns source clusters, $L_i$ for $i\in\{1,2,\ldots,Q\}$ where $Q\ll K$, on which the sky variation is considered to be uniform. Then, it assumes there exists a unique $\widetilde{{\bf{J}}}_{pi}$ which is shared by all the sources of the $i$-th cluster $L_i$, $i\in\{1,2,\ldots,Q\}$, at receiver $p$, $p\in\{1,2,\ldots,N\}$. Based on that, the visibility at every baseline $p-q$, given by Eq. (\ref{ss3}), is reformulated as
\begin{equation}
\widetilde{\bf V}_{pq}=\sum_{i=1}^Q \widetilde{{\bf{J}}}_{pi}({\widetilde{\pmb{\theta}}})\{\sum_{l\in L_i}{}{{\bf{C}}}_{l\{pq\}}\}\widetilde{{\bf{J}}}^H_{qi}({\widetilde{\pmb{\theta}}})+\widetilde{\bf{N}}_{pq}.\label{sa3}
\end{equation}
In Eq. (\ref{sa3}), $\widetilde{\bf{N}}_{pq}$ is the clustered calibration's effective noise at baseline $p-q$ which will be explicitly discussed at section \ref{Performance Analysis}. Note that clustered calibration estimates the new unknown parameter $\widetilde{\pmb{\theta}}\in \mathbb{C}^{\widetilde{P}}$ where $\widetilde{P}=4QN$. Denoting the effective signal of every $i$-th cluster at baseline $p-q$ by
\begin{equation}
\widetilde{{\bf{C}}}_{i\{pq\}}\equiv\sum_{l\in L_i}{}{{\bf{C}}}_{l\{pq\}},\label{sa2}
\end{equation}
the clustered calibration visibility vector at this baseline (vectorized form of Eq. (\ref{sa3})) is 
\begin{equation}
{\bf v}_{pq}=\sum_{i=1}^Q \widetilde{\bf J}^*_{qi}(\widetilde{\pmb{\theta}})\otimes\widetilde{\bf J}_{pi}(\widetilde{\pmb{\theta}})\mbox{vec}(\widetilde{\bf C}_{i\{pq\}})+\widetilde{\bf n}_{pq}.\label{s33}
\end{equation}
Finally, stacking up the visibilities of all the instrument's baselines in vector ${\bf y}$, clustered calibration's general measurement equation is resulted as
\begin{equation}
{\bf y}=\sum_{i=1}^Q \widetilde{\bf s}_i(\widetilde{\pmb{\theta}})+\widetilde{\bf n}.\label{sa4}
\end{equation}
In Eq. (\ref{sa4}) $\widetilde{\bf s}_i$ is defined similar to ${\bf s}_i$ in Eq. (\ref{sa5}) where ${\bf J}$ and ${\bf C}$ are replaced by $\widetilde{\bf J}$ and $\widetilde{\bf C}$, respectively. 

Because of the similarity between the clustered and the un-clustered calibration's measurement equations presented by Eq. (\ref{sa4}) and Eq. (\ref{s4}), respectively, they could utilize the same calibration techniques. Thus, the only difference between these two types of calibration is that clustered calibration solves for clusters of sources instead of for the individual ones. That upgrades the signals which should be calibrated for from Eq. (\ref{sa22}) to Eq. (\ref{sa2}).

\section{Clustering Algorithms}\label{clustering}
Clustering is grouping a set of data so that the members of the same group (cluster) have some similarities \citep{Clustering}. This similarity is defined based on the application of the clustering method.

We need to define clustering schemes in which two radio sources merge to a single cluster based on the similarity in their direction dependent gain errors ( Eq. (\ref{i3}) ). Radiations of sources that are close enough to each other in the sky are assumed to be affected by almost the same corruptions ( Eq. (\ref{i4}) ). Based on that, we aim to design source clusters with small angular diameters. On the other hand, every cluster's intensity is the sum of the intensities of its members ( Eq. (\ref{sa2}) ). In order to keep a balance between different clusters' intensities, we intend to apply weighted clustering techniques in which the sources are weighted proportional to their intensities. 

Suppose that the $K$ sources, $x_1,x_2,\ldots,x_K$ have $(\alpha_1,\delta_1),(\alpha_2,\delta_2),\ldots,(\alpha_K,\delta_K)$ equatorial coordinates, respectively. The aim is to provide $Q$ clusters so that the objective function $f=\sum_{q=1}^Q D(L_q)$ is minimized. $D(L_q)$ is the angular diameter of cluster $L_q$, for $q\in\{1,2,\ldots,Q\}$, defined as
\hspace*{-6mm}\begin{equation}
D(L_q)\equiv \mbox{max}\ \{d(x_i,x_j)|x_i,x_j\in L_q\},
\end{equation}
and $d(.,.)$ is the angular separation between any two points on the celestial sphere. Having two radio sources $a$ and $b$ with equatorial coordinates $(\alpha_a,\delta_a)$ and $(\alpha_b,\delta_b)$, respectively, the angular separation $d(a,b)$, in radians, is obtained by
\begin{equation}
\mbox{tan}^{-1}\frac{\sqrt{\mbox{cos}^2\delta_b\mbox{sin}^2\Delta\alpha+[\mbox{cos}\delta_a\mbox{sin}\delta_b-\mbox{sin}\delta_a\mbox{cos}\delta_b\mbox{cos}\Delta\alpha]^2}}{\mbox{sin}\delta_a\mbox{sin}\delta_b+\mbox{cos}\delta_a\mbox{cos}\delta_b\mbox{cos}\Delta\alpha},
\end{equation}\label{sp1}
where $\Delta\alpha=\alpha_b-\alpha_a$.\\
For defining the centroids, we associate a weight to every source $x_i$, as
\begin{equation}
w_i=w(x_i)\equiv \frac{I_i}{I^*},\quad\mbox{for}\ i\in\{1,2,\ldots,K\},
\end{equation}\label{Eq.2}
where $I_i$ is the source's intensity and $I^*=\mbox{min}\ \{I_1,I_2,\ldots,I_K\}$. Applying the weights to the clustering procedure, the centroids of the clusters lean mostly towards the brightest sources. That causes a tendency in faint sources to gather with brighter sources close to them into one cluster. Thus, their weak signals are promoted being added up with some brighter sources' signals. Moreover, very strong sources will be isolated such that their signals are calibrated individually, without being affected by the other faint sources. 

We cluster radio sources using weighted K-means \citep{weightedk-means} and divisive hierarchical clustering \citep{Hierarchicalclustering} algorithms. Since the source clustering for calibration is performed offline, its computational complexity is negligible compared with the calibration procedure itself. 

\subsection{Weighted K-means clustering}\label{Weighted K-means clustering}
{\bf Step 1.} Select the $Q$ brightest sources, $x_{1^*},x_{2^*},\ldots,x_{Q^*}$, and initialize the centroids of $Q$ clusters by their locations as
\begin{equation}
c_q\equiv [\alpha_{q^*},\delta_{q^*}],\quad \mbox{for}\ q\in\{1,2,\ldots,Q\},\ q^*\in\{1^*,2^*,\ldots,Q^*\}.
\end{equation}
{\bf Step 2.} Assign each source to the cluster with the closest centroid, defining the membership function
\begin{equation*}
m_{L_q}(x_i)=\begin{cases} 1, & \mbox{if }\  d(x_i,c_q)=\mbox{min}\{d(x_i,c_j)|j=1,\ldots,Q\}  \\ 0,  & \mbox{Otherwise } \end{cases}
\end{equation*}
{\bf Step 3.} Update the centroids by
\begin{equation}
c_q=\frac{\sum_{i=1}^Km_{L_q}(x_i)\ w_ix_i}{\sum_{i=1}^Km_{L_q}(x_i)\ w_i},\quad \mbox{for}\ q\in\{1,2,\ldots,Q\}.
\end{equation}
Repeat steps 2 and 3 until there are no reassignments of sources to clusters.

\subsection{Divisive hierarchical clustering}
\label{Divisive hierarchical clustering algorithm}
{\bf Step 1.} Initialize the cluster counter $Q'$ to $1$, assign all the $K$ sources to a single cluster $L_1$ and $\varnothing$ to a set of null clusters $A$.\\
{\bf Step 2.} Choose cluster $L_{q^*}$, for $q^*\in\{1,2,\ldots,Q'\}-A$, with the largest angular diameter
\begin{equation}
D(L_{q^*})=\mbox{max}\{D(L_{q})|q\in\{1,2,\ldots,Q'\}-A\}.
\end{equation}
{\bf Step 3.} Apply the presented weighted K-means clustering technique to split $L_{q^*}$ into two clusters, $L'_{q^*}$ and $L''_{q^*}$.\\\\
{\bf Step 4.} If $D(L'_{q^*})+D(L''_{q^*})<D(L_{q^*})$, then set $Q'=Q'+1$, $L_{q^*}\equiv L'_{q^*}$, $L_{Q'}\equiv L''_{q^*}$, and $A=\varnothing$, otherwise set $A=A\cup\{q^*\}$.\\
Repeat steps 2, 3, and 4 until $Q'=Q$.

\subsection{Comparison of clustering methods}
\label{Clustering methods comparison}
Hierarchical clustering method tends to design clusters with almost the same angular diameters, whereas, the K-means clustering method tends to keep the same level of intensity at all its clusters. In practice, since hierarchical clustering method makes less errors in dedicating the same solutions to sources in small clusters, it performs better than Weighted K-means clustering in a clustered calibration procedure. But, when the number of sources (and clusters) is very large ($Q\geq100$), its prohibitive computational costs makes the fast K-means clustering method preferable. 

\subsubsection{Example 1: Weighted K-means and hierarchical clustering}
\begin{figure}
\centering
\scalebox{0.25}{\includegraphics*[-110,-15][710,780]{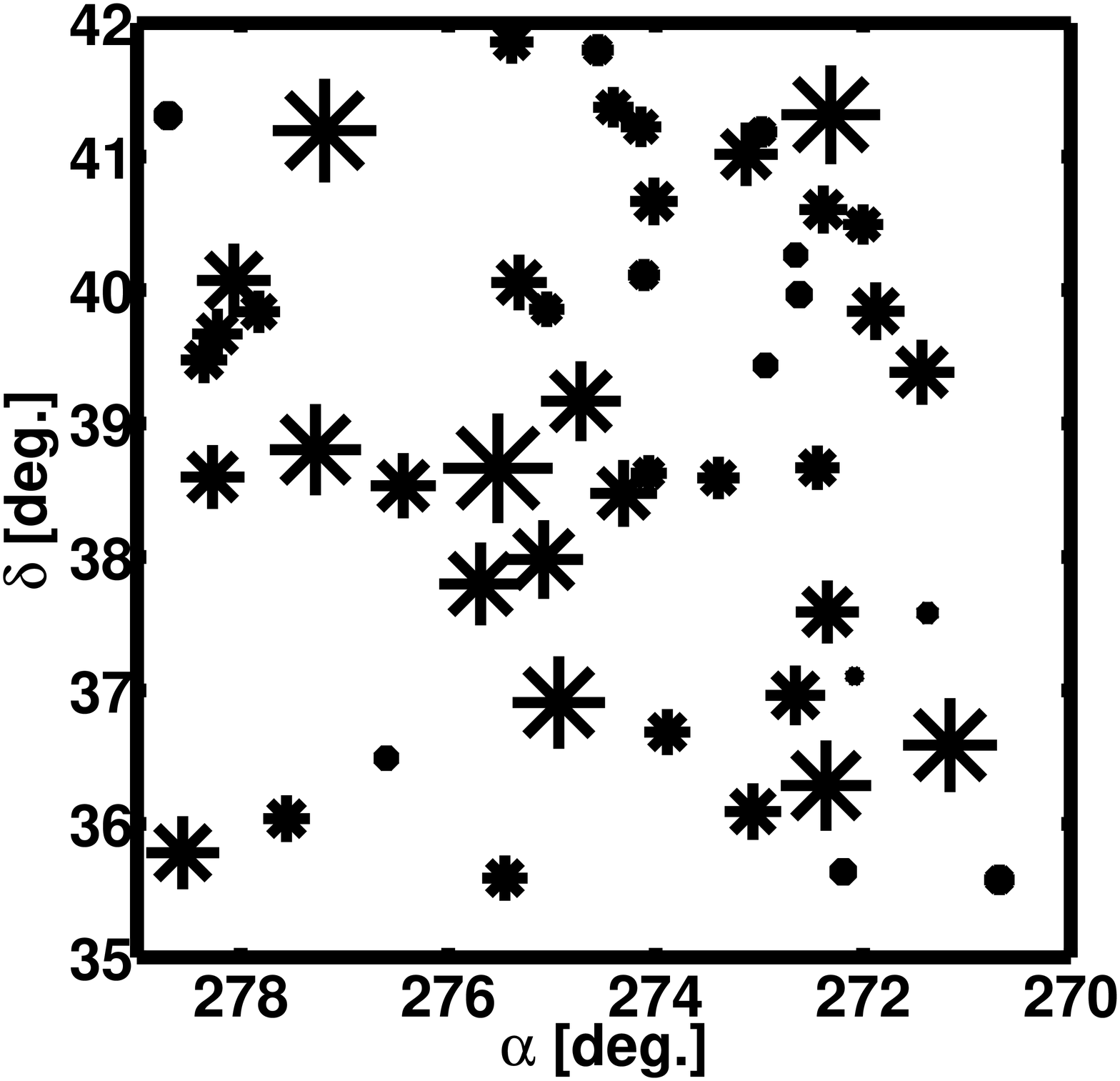}}
\caption{A simulated 8 by 8 degrees sky of fifty point sources with intensities below 3 Jy. The source positions and their brightness are following uniform and Rayleigh distributions, respectively. The marker sizes are proportional to sources intensities.}
\label{figure15}
\end{figure}

\begin{figure}
\centering
\scalebox{0.25}{\includegraphics*[-110,-15][710,780]{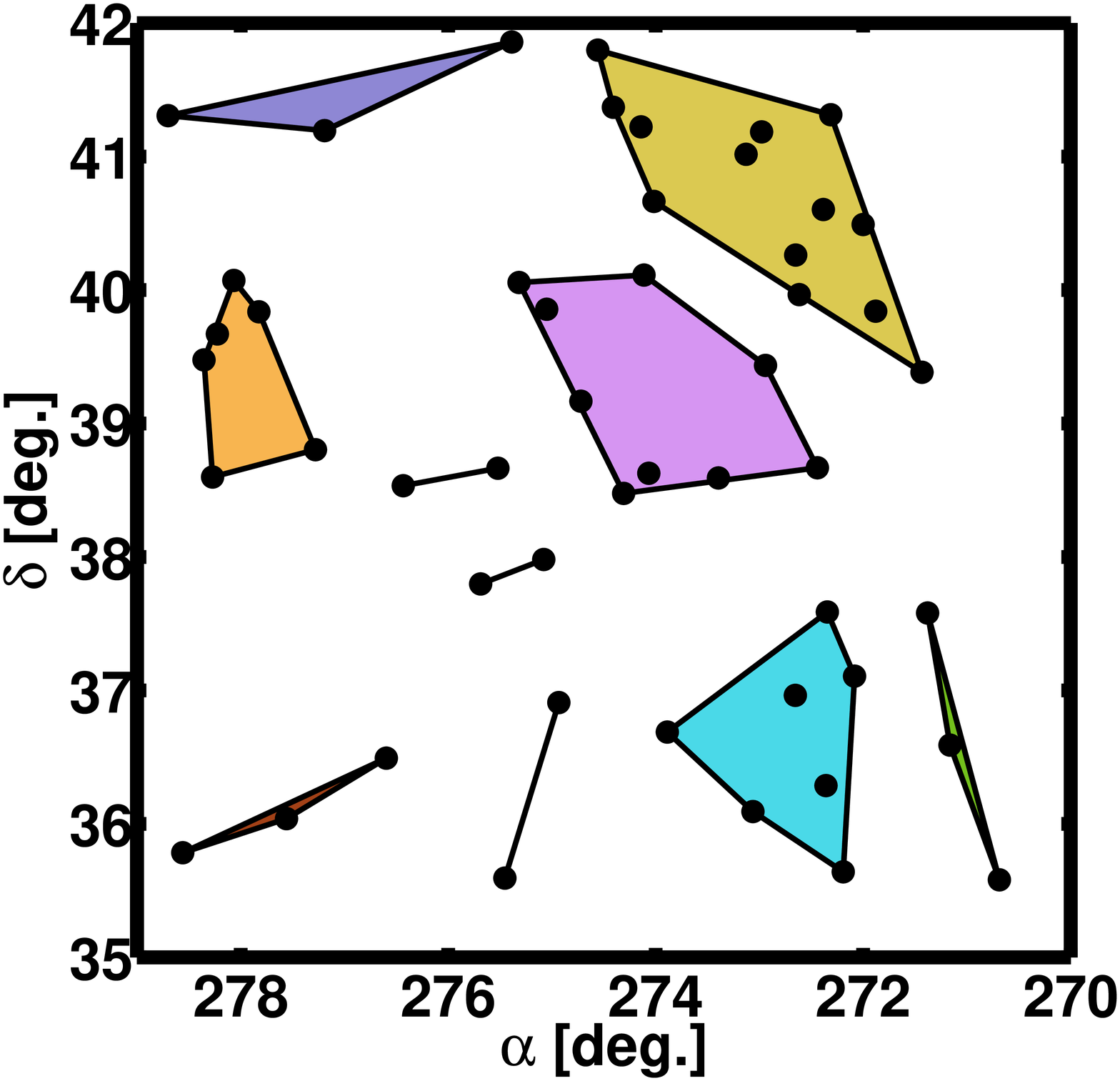}}
\caption{Fifty point sources are clustered into ten source clusters by Weighted K-means clustering technique. There is not a good balance between different clusters angular diameters.}
\label{figure14}
\end{figure}

\begin{figure}
\centering
\scalebox{0.25}{\includegraphics*[-110,-15][710,780]{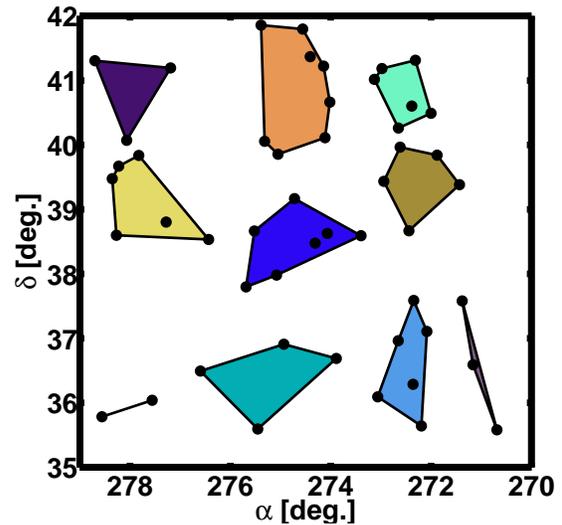}}
\caption{Fifty point sources are clustered into ten source clusters via hierarchical clustering method. Different clusters have almost the same angular diameters. }
\label{figure13}
\end{figure}

We simulate an 8 by 8 degrees sky with fifty point sources with intensities below 3 Jansky (Jy).  The source positions and their brightness follow uniform and Rayleigh distributions, respectively. The result is shown by Fig. \ref{figure15} in which the symbol sizes are proportional to intensities of sources.   
Weighted K-means and divisive hierarchical clustering methods are applied to cluster the fifty sources into ten source clusters. The results are presented in Fig. \ref{figure14} and Fig. \ref{figure13}, respectively. Fig. \ref{figure14} shows that the Weighted K-means clustering could design source clusters with considerably large angular diameters. Assigning the same calibration solutions to the sources of these large clusters could cause significant errors. However, as Fig. \ref{figure13} shows, this is not the case for the hierarchical clustering and it constructs clusters with almost the same angular diameters. 

\begin{figure}
\centering
\vspace{-0.0cm}
\hspace*{-5mm}
\includegraphics[width=9.5cm,height=6cm]{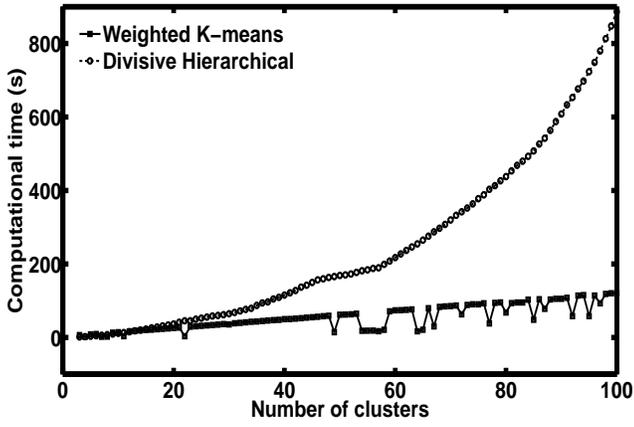}
\vspace{-0.5cm}
\caption{Weighted K-means and divisive hierarchical clustering methods computational costs. For small number of source clusters, there is no difference between execution times of the two clustering methods. But, when the number of source clusters is large, the computational cost of Weighted K-means becomes much cheaper than the one of the hierarchical clustering. }
\label{figure16}
\end{figure}

Since the number of sources in this simulation is not that large ($K=50$), the difference between execution time of the two clustering methods is not significant. Hence in such a case, the use of hierarchical clustering method, rather than the Weighted K-means, is advised. However, this is not the case when we have a large number of sources, and subsequently a large number of source clusters, in the sky. To demonstrate this, we use the two clustering techniques for clustering thousand of sources ($K=1000$) into $Q$ source clusters, $Q\in\{3,4,\ldots,100\}$. The methods' computational times versus the number of clusters are plotted in Fig. \ref{figure16}. As Fig. \ref{figure16} shows, for large $Q$s, the computational cost of Weighted K-means is much cheaper than the one of the hierarchical clustering. That can make the Weighted K-means clustering method more suitable than the hierarchical clustering for such a case.

\section{Performance Analysis}\label{Performance Analysis}
 In this section, we explain the reasons for clustered calibration's better performance, compared to un-clustered calibration, at a low SNR \citep{S.K.3}. This superiority is in the sense of achieving an unprecedented precision in solutions with a considerably low computational complexity, given the optimum clustering scheme. In the next section, we present different criteria for finding the optimum number of clusters at which the clustered calibration performs the best.

\subsection{Cramer-Rao Lower Bounds}
The most fundamental assumption in clustered calibration is that the sources at the same cluster have exactly the same corruptions in their radiated signals. This assumption is of course incorrect, nonetheless, it provides us with a stronger signal, the sum of the signals in the whole cluster. We present an analytic  comparison of clustered and un-clustered calibration where we use the Cramer-Rao Lower Bound (CRLB) \citep{CRVB} as a tool to measure the  performance of the calibration.

\subsubsection{Estimations of CRLB for two sources at a single cluster}\label{Estimations of Cramer-Rao Lower Bounds}
For simplicity, first consider observing two point sources at a single baseline, lets say baseline $p-q$. Based on Eq. (\ref{ss3}), the visibilities are given by
\begin{equation}
\widetilde{\bf V}_{pq}={\bf J}_{p1}{\bf C}_{1\{pq\}}{\bf J}_{q1}^H + {\bf J}_{p2}{\bf C}_{2\{pq\}}{\bf J}_{q2}^H +{\bf N}_{pq},\label{s5}
\end{equation}
in the un-clustered calibration strategy. Vectorizing $\widetilde{\bf V}_{pq}$, the visibility vector is
\begin{equation*}
{\bf y}={\bf J}^*_{q1}\otimes{\bf J}_{p1}\mbox{vec}({\bf C}_{1\{pq\}})+{\bf J}^*_{q2}\otimes{\bf J}_{p2}\mbox{vec}({\bf C}_{2\{pq\}})+{\bf n}_{pq}.
\end{equation*} 
Assuming ${\bf n}_{pq}\sim\mathcal{CN}({\bf 0},\sigma^2{\bf I}_4)$, we have 
\begin{equation}
{\bf y}\sim \mathcal{CN}({\bf s}({\pmb {\theta}}),\sigma^2{\bf I}_4), \label{d1}
\end{equation}
where
\begin{equation}
{\bf s}({\pmb {\theta}})\equiv\sum_{i=1,2}{\bf J}^*_{qi}({\pmb {\theta}})\otimes{\bf J}_{pi}({\pmb {\theta}})\mbox{vec}({\bf C}_{i\{pq\}}).
\end{equation}
Using Eq. (\ref{d1}), the log-likelihood function of the visibility vector ${\bf y}$ is given by 
\begin{equation}
\mathcal{L}({\pmb {\theta}} |{\bf y})=-4\mbox{ln}\{\frac\pi{\sigma^2}\}-{\sigma^{-2}}({\bf y}-{\bf s}({\pmb {\theta}}))^H({\bf y}-{\bf s}({\pmb {\theta}})).\label{crb4}
\end{equation}

CRLB is a tight lower bound on the error variance of any unbiased parameter estimators \citep{CRVB}. Based on its definition, if the log-likelihood function of the random vector ${\bf y}$, $\mathcal{L}({\pmb {\theta}} |{\bf y})$, satisfies the ``regularity'' condition 
\begin{equation}
\operatorname{E}_{\bf y} \left[ \frac{\partial}{\partial{\pmb{\theta}}} \mathcal{L}({\pmb {\theta}} |{\bf y}) \right]=0,\quad\mbox{for all}\ {\pmb{\theta}},
\end{equation}
then for any unbiased estimator of ${\pmb{\theta}}$, $\widehat{\pmb{\theta}}$,
\begin{equation}
\mbox{var}(\widehat{\pmb{\theta}}_i)\geq[\mathcal{I}^{-1}({\pmb{\theta}})]_{ii},\quad\mbox{for}\  i\in\{1,\ldots,M\},\label{q2}
\end{equation}
where $\mathcal{I}({\pmb{\theta}})$ is the Fisher information matrix defined as
\begin{equation}
\mathcal{I}({\pmb{\theta}})=- \operatorname{E}_{\bf y} \left[ \frac{\partial^2 \mathcal{L}({\pmb {\theta}} |{\bf y})}{\partial{\pmb{\theta}} \partial{\pmb{\theta}}^T} \right].\label{news}
\end{equation}
In other words, the variance of any unbiased estimator of the unknown parameter vector ${\pmb{\theta}}$ is bounded from below by the diagonal elements of $[\mathcal{I}({\pmb{\theta}})]^{-1}$.

Using (\ref{crb4}) and (\ref{news}), the Fisher information matrix of the visibility vector ${\bf y}$ is obtained as
\begin{equation}
\mathcal{I}({\pmb {\theta}})=2\sigma^{-2}\ \mathfrak{Re}(J_{\bf s}^HJ_{\bf s}),
\end{equation}
where $J_{\bf s}$ is the Jacobian matrix of ${\bf s}$ with respect to ${\pmb {\theta}}$
\begin{equation}
J_{\bf s}({\pmb \theta})=\sum_{i=1}^2 {\frac\partial {\partial{\pmb{\theta}}}}\{{\bf J}_{qi}^*\otimes{\bf J}_{pi}\}[{\bf I}_4\otimes\mbox{vec}({\bf C}_{i\{pq\}})].\label{khoda}
\end{equation}
Thus, variations of any unbiased estimator of parameter vector ${\pmb {\theta}}$, lets say $\widehat{\pmb {\theta}}$, is bounded from below by the CRLB as
\begin{equation}
\mbox{Var}(\widehat{\pmb {\theta}})\geq[2\sigma^{-2}\ \mathfrak{Re}(J_{\bf s}^HJ_{\bf s})]^{-1}.\label{crb9}
\end{equation}

Lets try to bound the error variations of the clustered calibration parameters assuming that the two sources construct a single cluster, called cluster number $1$. We reform Eq. (\ref{s5}) as
\begin{equation}
{\bf V}_{pq}=\widetilde{\bf J}_{p1}({\bf C}_{1\{pq\}}+{\bf C}_{2\{pq\}})\widetilde{\bf J}_{q1}^H + {\bf \Gamma}_{1\{pq\}}+ {\bf \Gamma}_{2\{pq\}}+{\bf N}_{pq},\label{s6}
\end{equation}
where ${\bf \Gamma}_{i\{pq\}}$, referred to as the ``clustering error" matrices, are given by
\begin{equation}
{\bf \Gamma}_{i\{pq\}}={\bf J}_{pi}{\bf C}_{i\{pq\}}{\bf J}_{qi}^H - \widetilde{\bf J}_{p1}{\bf C}_{i\{pq\}}\widetilde{\bf J}_{q1}^H,\label{s18}
\end{equation}
and $\widetilde{\bf J}_{p1}(\widetilde{\pmb {\theta}})$ is the clustered calibration solution at receiver $p$.\\
Eq. (\ref{s6}) implies that what is considered as the noise matrix $\widetilde{\bf N}_{pq}$ in the clustered calibration data model, Eq. (\ref{sa3}), is in fact
\begin{equation}
\widetilde{\bf N}_{pq}\equiv{\bf \Gamma}_{1\{pq\}}+ {\bf \Gamma}_{2\{pq\}}+{\bf N}_{pq}.\label{s7}
\end{equation}
Vectorizing Eq. (\ref{s6}), the clustered calibration visibility vector is obtained by
\begin{equation}
{\bf y}=\widetilde{\bf J}^*_{q1}\otimes\widetilde{\bf J}_{p1}\mbox{vec}({\bf C}_{1\{pq\}}+{\bf C}_{2\{pq\}})+\widetilde{\bf n}_{pq},\label{d2}
\end{equation} 
where $\widetilde{\bf n}_{pq}=\mbox{vec}(\widetilde{\bf N}_{pq})$.

We point out that depending on the observation as well as the positions of the two sources on the sky, the clustering error ${\bf \Gamma}_{i\{pq\}}$ will have different properties. However, in order to study the performance of the clustered calibration in  a statistical sense, and to simplify our analysis, we make the following assumptions.
\begin{enumerate}
\item Consider statistical expectation over different observations and over different sky realizations where the sources are randomly distributed on the sky. In that case, almost surely $\operatorname{E}\{\widetilde{\bf J}\}\rightarrow \operatorname{E}\{{\bf J}\}$ and consequently 
\begin{equation}
\operatorname{E}\{{\bf \Gamma}_{i\{pq\}}\}\rightarrow {\bf 0}.\label{s8}
\end{equation}
In other words, we assume the clustering error to have zero mean over many observations of different parts of the sky.
\item We assume that the closer the sources are together in the sky, the smaller the errors  introduced by clustering would be. Therefore, given a set of sources, the clustering error will reduce as the number of clusters increase. In fact this error introduced by clustering vanishes when the number of clusters is equal to the number of sources (each cluster contains only one source). Therefore, given a set of sources, the variance of ${\bf \Gamma}_{i\{pq\}}$  will decrease as the number of clusters increase.
\end{enumerate}

Using Eq. (\ref{s8}) and bearing in mind that $\operatorname{E}\{{\bf N}_{pq}\}={\bf 0}$, we can consider $\widetilde{\bf n}_{pq}\sim\mathcal{CN}({\bf 0},\widetilde{\sigma}^2{\bf I}_4)$ where $\operatorname{E}\{\widetilde{\bf n}_{pq}\widetilde{\bf n}^H_{pq}\}=\widetilde{\sigma}^2{\bf I}_4$. Therefore,
\begin{equation*}
{\bf y}\sim \mathcal{CN}(\widetilde{\bf s},\widetilde{\sigma}^2{\bf I}_4),
\end{equation*}
\begin{equation*}
\widetilde{\bf s}\equiv\widetilde{\bf J}^*_{q1}\otimes\widetilde{\bf J}_{p1}\mbox{vec}({\bf C}_{1\{pq\}}+{\bf C}_{2\{pq\}}),
\end{equation*}
and similar to Eq. (\ref{crb9}), we have
\begin{equation}
\mbox{Var}(\widehat{\widetilde{\pmb {\theta}}})\geq[2\widetilde{\sigma}^{-2}\ \mathfrak{Re}(J_{\widetilde{\bf s}}^HJ_{\widetilde{\bf s}})]^{-1}.\label{d10}
\end{equation}
We use numerical simulations to compare the un-clustered and clustered calibrations performances via their CRLBs which are given by Eq. (\ref{crb9}) and Eq. (\ref{d10}), respectively.

\subsubsection{Example 2: CRLB for two sources and one cluster}
\label{Simulation 1}
\begin{figure}
\centering
\vspace{-0.0cm}
\hspace*{-5mm}
\includegraphics[width=9.5cm,height=6cm]{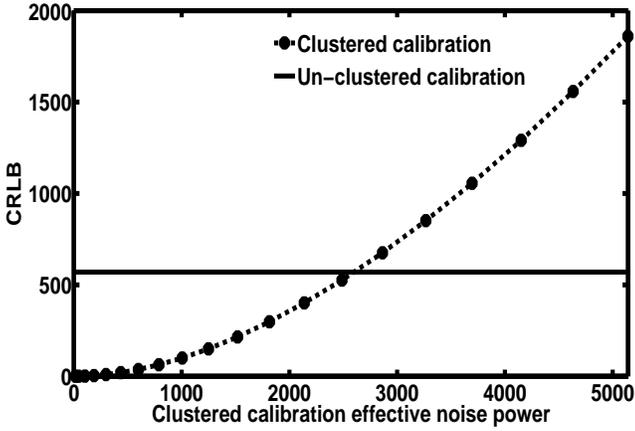}
\vspace{-0.5cm}
\caption{Clustered and un-clustered calibrations CRLB. When the effective noise power of the clustered calibration, $||\widetilde{\bf N}||^2$, is small enough, then its CRLB is lower than of the un-clustered calibration's and it reveals a superior performance. }
\label{figure1}
\end{figure}

We simulate a twelve hour observation of two point sources with intensities $I_1=11.25$ and $I_2=2.01$ Jy at sky coordinates $(l,m)$ equal to  $(-0.014,-0.005)$ and $(-0.011,-0.010)$ radians, respectively. We use the uv-coverage of Westerbork Synthesis Radio Telescope (WSRT) with 14 receivers in this simulation.\\
We Consider the ${\bf J}$ Jones matrices in Eq. (\ref{s5}) to be diagonal. Their amplitude and phase elements follow $\mathcal{U}(0.75,0.95)$ and $\mathcal{U}(0.003,0.004)$ distributions, respectively. The background noise is ${\bf N}\sim\mathcal{CN}({\bf 0},10{\bf I})$. Jones matrices of the clustered calibration, $\widetilde{\bf J}_{p1}$ for $p=1,2$, are obtained as ${\bf J}_{p1}+\mathcal{U}(0.02,0.40)e^{j\mathcal{U}(0.5,2)}$. For 20 realizations of $\widetilde{\bf J}$ matrices, we calculated CRLB of the un-clustered and clustered calibrations using Eq. (\ref{crb9}) and Eq. (\ref{d10}), respectively. The results are presented in Fig. \ref{figure1}. As shown in this figure, for small enough errors matrices ${\pmb \Gamma}$ of Eq. (\ref{s18}), the clustered calibration's CRLB stands below the un-clustered calibration's CRLB. On the other hand, with increasing power of error matrices, or the power of effective noise $\widetilde{\bf N}$, the un-clustered calibration's CRLB becomes lower than the clustered calibration's CRLB. 

\subsubsection{Analysis of CRLB}
\label{Analysis}
Generally, if source $1$ is considerably brighter than source $2$, $||{\bf C}_{1\{pq\}}|| \gg ||{\bf C}_{2\{pq\}}||$, and if the weak source power is much lower than the noise level, $||{\bf C}_{2\{pq\}}|| \ll ||{\bf N}_{pq}||$, then clustered calibration's performance is better than un-clustered calibration. Note that the worst performance of both calibrations is at the faintest source and we are more concerned to compare the CRLBs for this source. 

The CRLBs obtained for the un-clustered and clustered calibrations in Eq. (\ref{crb9}) and Eq. (\ref{d10}), respectively, are both almost equal to the inverse of the Signal to Interference plus Noise Ratio (SINR), $\mbox{SINR}^{-1}$. In un-clustered calibration, the effective signal for the faintest source is ${\bf C}_{2\{pq\}}$ where the noise is ${\bf N}_{pq}$. Therefore, SINR for this source is 
\begin{equation}
\mbox{SINR}_2=\frac{||{\bf C}_{2\{pq\}}||^2}{||{\bf N}_{pq}||^2}.
\end{equation}
But, in clustered calibration, the effective signal and noise are $\widetilde{\bf C}_{\{pq\}}\equiv{\bf C}_{1\{pq\}}+{\bf C}_{2\{pq\}}$ and $\widetilde{\bf N}_{pq}$, respectively. Thus, SINR for the cluster is
\begin{equation}
\mbox{SINR}_c=\frac{||\widetilde{\bf C}_{\{pq\}}||^2}{||\widetilde{\bf N}_{pq}||^2}.
\end{equation}
Clustered calibration has an improved performance when 
\begin{equation}
\mbox{SINR}_c\gg\mbox{SINR}_2.\label{s21}
\end{equation}

Consider the two possible extremes in a clustered calibration procedure: 
\begin{enumerate}
\item{ Clustering many sources in a large field of view to a very small number of clusters. In this case, the angular diameter of a cluster is probably too large for the assumption of uniform corruptions to apply. Subsequently, dedicating a single solution to all the sources of every cluster by clustered calibration introduces clustering error matrices ${\pmb \Gamma}$ with a large variance (see Eq. (\ref{s18})). Having high interference power, the clustered calibration effective noise $\widetilde{\bf N}$ of Eq. (\ref{s7}) becomes very large. Therefore, clustered calibration SINR will be very low and it does not produce high quality results. }
\item{ Clustering sources in a small field of view to a very large number of clusters. In this case, the variance of ${\pmb \Gamma}$ matrices are almost zero while the signal powers of source clusters are almost as low as of the individual sources. Therefore, the SINR of clustered calibration is almost equal to the un-clustered calibration SINR and the calibration performance is expected to be almost the same as well. \\
Thus, the  best efficiency of clustered calibration is obtained at the smallest number of clusters for which Eq. (\ref{s21}) is satisfied. We use the SINR of Eq. (\ref{s21}) as an efficient criterion for detecting the optimum number of clusters. }
\end{enumerate}

\subsubsection{Generalization to many sources and many clusters}
\label{Generalizations}
For the visibility vector ${\bf y}$ of un-clustered calibration's general data model, presented by Eq. (\ref{s4}), we have
\begin{equation}
{\bf y}\sim \mathcal{CN}(\sum_{i=1}^{K} {\bf s}_i({\pmb {\theta}}),{\pmb{\Pi}}).\label{i1}
\end{equation}
Therefore, the CRLB of un-clustered calibration is
\begin{equation*}
\mbox{var}({\pmb{\theta}})\geq\bigg[2\ \mathfrak{Re}\bigg\{(\sum_{i=1}^{K}J_{{\bf s}_i}({\pmb{\theta}}))^H {\pmb \Pi}^{-1} (\sum_{i=1}^{K}J_{{\bf s}_i}({\pmb{\theta}}))\bigg\}\bigg]^{-1},\label{f7}
\end{equation*}
where $J_{{\bf s}_i}$ is the Jacobian matrix of ${{\bf s}_i}$ with respect to ${\pmb {\theta}}$.

Computing the exact CRLB is more complicated when we have clustered calibration. In the clustered calibration measurement equation, given by Eq. (\ref{sa4}), we have
\begin{equation}
\widetilde{\bf n}\equiv\sum_{i=1}^K {\pmb \Gamma}_i+{\bf n},\label{nn4}
\end{equation}
 where ${\bf n}$ is the un-clustered calibration's noise vector,
\begin{equation}
{\pmb \Gamma}_i=[\mbox{vec}({\pmb \Gamma}_{i\{12\}})^T\ldots \mbox{vec}({\pmb \Gamma}_{i\{(N-1)N\}})^T]^T,\label{s16}
\end{equation}
and ${\pmb \Gamma}_{i\{pq\}}$ is given by Eq. (\ref{s18}). Due to the existence of the nuisance parameters ${\pmb \Gamma}_i$ in the clustered calibration data model, calculation of its conventional CRLB is impractical. This leads us to the use of Cramer-Rao like bounds devised in the presence of the nuisance parameters \citep{CRLB}. We apply the Modified CRLB (MCRLB) \citep{MCRVB2} to the performance of clustered calibration.

The MCRLB for estimating the errors of $\widehat{\tilde{\pmb {\theta}}}$ in the presence of the nuisance parameters ${\pmb \Gamma}$ (clustering error) is defined as 
\begin{equation}
\mbox{var}(\widehat{\tilde{\pmb {\theta}}})\geq\bigg[\operatorname{E}_{{\bf y},{\pmb \Gamma}} \bigg\{-\operatorname{E}_{{\bf y}|{\pmb \Gamma}} \bigg\{\frac{\partial}{\partial{\tilde{\pmb{\theta}}}} \frac{\partial}{\partial{\tilde{\pmb{\theta}}}^T}\mbox{ln}\{P({\bf y}|{\pmb \Gamma};\tilde{\pmb{\theta}})\} \bigg\}\bigg\}\bigg]^{-1},\label{f5}
\end{equation}
where  $P({\bf y}|{\pmb \Gamma};\tilde{\pmb{\theta}})$ is the Probability Density Function (PDF) of the visibility vector ${\bf y}$ assuming that the ${\pmb \Gamma}$ matrices of Eq. (\ref{s16}) are a priori known. Since ${\bf n}\sim \mathcal{CN}({\bf 0},{\pmb{\Pi}})$, from Eq. (\ref{sa4}) we have 
\begin{equation}
{\bf y}|{\pmb \Gamma}\sim \mathcal{CN}([\sum_{i=1}^Q \tilde{\bf s}_i+\sum_{i=1}^K{\pmb \Gamma}_i],{\pmb{\Pi}}),\label{f4}
\end{equation}
and therefore in Eq. (\ref{f5}), $-\operatorname{E}_{{\bf y}|{\pmb \Gamma}} \bigg[\frac{\partial}{\partial{\tilde{\pmb{\theta}}}} \frac{\partial}{\partial{\tilde{\pmb{\theta}}}^T} \mbox{ln}\{P({\bf y}|{\pmb \Gamma};\tilde{\pmb{\theta}})\}\bigg]$, which is called the modified Fisher information, is equal to
\begin{equation*}
2\mathfrak{Re}\{[\sum_{i=1}^{Q}J_{\tilde{\bf s}_i}(\tilde{\pmb{\theta}})+\sum_{i=1}^{K}J_{{\pmb \Gamma}_i}(\tilde{\pmb{\theta}})]^H {\pmb \Pi}^{-1} [\sum_{i=1}^{Q}J_{\tilde{\bf s}_i}(\tilde{\pmb{\theta}})+\sum_{i=1}^{K}J_{{\pmb \Gamma}_i}(\tilde{\pmb{\theta}})]\}.
\end{equation*}
 Note that $\operatorname{E}_{{\bf y},{\pmb \Gamma}}$ in Eq. (\ref{f5}) could be estimated by Monte-Carlo method.

As a rule of thumb, reducing the heavy computational cost of MCRLB, one can interpret the SINR test of Eq. (\ref{s21}) as follows: If in average the effective SINR of clustered calibration, $\mbox{SINR}_c$, gets higher than the effective SINR of un-clustered calibration obtained for the weakest observed signal, $\mbox{SINR}_w$, 
\begin{equation}
\operatorname{E}\{\mbox{SINR}_c\}\gg\operatorname{E}\{\mbox{SINR}_w\},\label{s22}
\end{equation}
then clustered calibration can achieve a better results. In Eq. (\ref{s22}), the expectation is taken with respect to the noise ${\bf N}$, error matrices ${\pmb \Gamma}$, and all the baselines.

\subsubsection{Example 3: MCRLB and SINR estimations}
\label{Simulation 2}
We simulate WSRT including $N=14$ receivers which observe fifty sources with intensities below 15 Jy. The source positions and their brightness follow uniform and Rayleigh distributions, respectively. The background noise is ${\bf N}\sim\mathcal{CN}({\bf 0},15{\bf I}_M)$, where $M=2N(N-1)=364$. We cluster sources using divisive hierarchical clustering, into $Q$ number of clusters where $Q\in\{3,4,\ldots,50\}$. Clustered calibration's Jones matrices, $\widetilde{{\bf J}}$, are generated as $\mathcal{U}(0.9,1.1)e^{j\mathcal{U}(0,0.2)}$. Since for smaller number of clusters, we expect larger interference (errors) in clustered calibration's solutions, for every $Q$, we consider $\sum_{i=1}^{50}{\bf \Gamma}_i\sim\mathcal{CN}({\bf 0},\frac{150}{Q}{\bf I}_M)$. The choice of the complex Gaussian distribution for the error matrices ${\pmb \Gamma}$ is due to the central limit theorem and the assumptions made in section \ref{Estimations of Cramer-Rao Lower Bounds}. 

\begin{figure}
\centering
\vspace{-0.0cm}
\hspace*{-5mm}
\includegraphics[width=9.5cm,height=6cm]{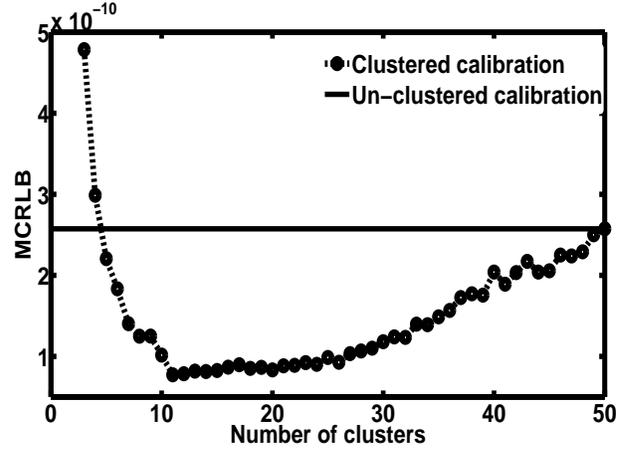}
\vspace{-0.5cm}
\caption{Clustered calibration's MCRLB. For very small $Q$, where the  effect of interference is large, MCRLB is high. By increasing the number of clusters, MCRLB decreases and reaches its minimum where the best performance of the clustered calibration is expected. After that, due to the dominant effect of the background noise, MCRLB starts to increase until it reaches the un-clustered calibration CRLB. }
\label{figure12}
\end{figure}

\begin{figure}
\centering
\vspace{-0.0cm}
\hspace*{-5mm}
\includegraphics[width=9.5cm,height=6cm]{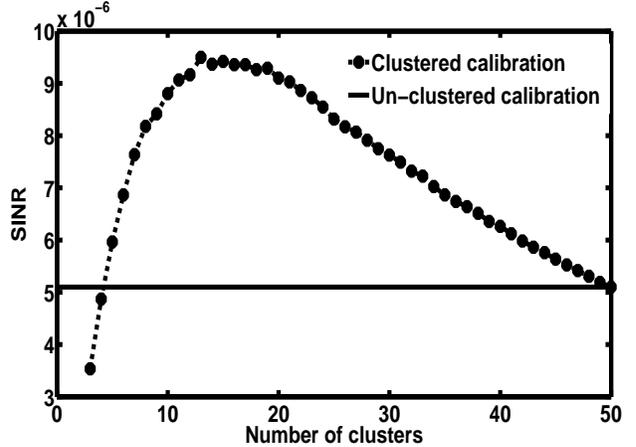}
\vspace{-0.5cm}
\caption{Clustered calibration's SINR. SINR is low for small $Q$, when the  interference is large. By increasing the number of clusters the SINR increases and gets its highest level for which the best performance of the calibration is expected. After that, it decreases due to the dominant effect of the background noise, and converges to the un-clustered calibration SINR. }
\label{figure2}
\end{figure}

We proceed to calculate the clustered calibration's MCRLB, given by Eq. (\ref{f5}), and $\operatorname{E}\{\mbox{SINR}_c\}$, utilizing the Monte-Carlo method. Jacobian matrices for MCRLB are calculated numerically and in computation of $\operatorname{E}\{\mbox{SINR}_c\}$, signal power of every cluster is obtained only using the cluster's brightest and faintest sources. The estimated results of MCRLB and $\operatorname{E}\{\mbox{SINR}_c\}$ are presented by Fig. \ref{figure12} and Fig. \ref{figure2}, respectively. As we can see in Fig. \ref{figure12}, for very small $Q$, where the  effect of interference is large, MCRLB is high. By increasing the number of clusters, MCRLB decreases and reaches its minimum where the best performance of the clustered calibration is expected. After that, due to the dominant effect of the background noise, MCRLB starts to increase until it reaches the CRLB of  un-clustered calibration. The same result is derived from $\operatorname{E}\{\mbox{SINR}_c\}$ plot of Fig. \ref{figure2}. As Fig. \ref{figure2} shows, $\operatorname{E}\{\mbox{SINR}_c\}$ is low for very small $Q$, when the interference (i.e., the error due to clustering) is large. By increasing the number of clusters, $\operatorname{E}\{\mbox{SINR}_c\}$ increases and gets its  peak for which the clustered calibration performs the best. After that, it decreases and converges to the  $\operatorname{E}\{\mbox{SINR}\}$ of un-clustered calibration. 


\subsection{Computational cost}
\label{computational aspects}
In the measurement equation of un-clustered calibration, presented in Eq. (\ref{s4}), we have $M=2N(N-1)$ constraints given by the visibility vector ${\bf y}$, and need to solve for $P=4KN$ unknown parameters ${\pmb{\theta}}$. If $P>M$, then Eq. (\ref{s4}) will be an under-determined non-linear system. This clarifies the need of having a small enough $N$ (number of antennas) and a large enough $K$ (number of sources) for estimating ${\pmb{\theta}}$. However, clustered calibration, Eq. (\ref{sa4}), has the advantage of decreasing the number of directions, $K$, relative to the number of source clusters, $Q\ll K$. This considerably cuts down the number of unknown parameters $P$ that needs to be calibrated, thus reducing the computational cost of calibration.

\section{Selection Of Number Of Clusters}\label{Model Order Selection}
Consider a clustered calibration procedure with a predefined clustering scheme. There is no guarantee that the calibration results for $Q$ number of clusters, where $Q\in\{1,2,\ldots,K\}$ is randomly chosen, is the most accurate. Thus, we seek the optimum number of clusters at which the clustered calibration performs the best. In this section, we describe the use of: (i) Akaike's Information Criterion (AIC) \citep{H.1}, as well as (ii) Likelihood-Ratio Test (LRT) \citep{L.R.T}, in finding this optimum $Q$ for a given observation. Some other alternative criteria could also be found in \citet{M.3}. Note that for different clustering schemes the optimum $Q$ is not necessarily the same.

\subsection{Akaike's Information Criterion (AIC)}
We utilize Akaike's Information Criterion (AIC) to find the optimum $Q$ for clustered calibration. \\
Consider having $\widetilde{\bf n}\sim\mathcal{CN}({\bf 0}, \widetilde{\sigma}^2{\bf I}_M)$ in the general data model of clustered calibration, Eq. (\ref{sa4}). Then, the log-likelihood of the visibility vector ${\bf y}$ is given by
\begin{eqnarray}
\mathcal{L}(\widetilde{\pmb {\theta}} |{\bf y})&=&-M\ \mbox{log}\ \pi -M\ \mbox{log}\ \widetilde{\sigma}^2 \nonumber \\
&-& \frac{1}{\widetilde{\sigma}^2}({\bf y}-\sum_{i=1}^Q\widetilde{\bf s}_i(\widetilde{\pmb {\theta}}))^H({\bf y}-\sum_{i=1}^Q\widetilde{\bf s}_i(\widetilde{\pmb {\theta}})).\label{aic1}
\end{eqnarray}
The maximum likelihood estimation of the noise variance $\widetilde{\sigma}^2$ is 
\begin{equation}
\widehat{\widetilde{\sigma}^2}=\frac{1}{M}({\bf y}-\sum_{i=1}^Q\widetilde{\bf s}_i(\widetilde{\pmb {\theta}}))^H({\bf y}-\sum_{i=1}^Q\widetilde{\bf s}_i(\widetilde{\pmb {\theta}})).\label{aic2}
\end{equation}
Substituting Eq. (\ref{aic2}) in Eq. (\ref{aic1}), we arrive at the maximum likelihood estimation of $\widetilde{\pmb {\theta}}$, 
\begin{eqnarray}
\mathcal{L}(\widehat{\widetilde{\pmb {\theta}}} |{\bf y})&=&-M\ \mbox{log}\ \pi -M\nonumber \\
&-& M \mbox{log}\{\frac{1}{M}({\bf y}-\sum_{i=1}^Q\widetilde{\bf s}_i(\widetilde{\pmb {\theta}}))^H({\bf y}-\sum_{i=1}^Q\widetilde{\bf s}_i(\widetilde{\pmb {\theta}}))\}.\label{aic3}
\end{eqnarray}
Using Eq. (\ref{aic3}), the AIC is given by
\begin{equation}
\mbox{AIC}(Q)=-2\mathcal{L}(\widehat{\widetilde{\pmb {\theta}}} |{\bf y})+2(2\widetilde{P}).\label{aic4}
\end{equation}
The optimum $Q$ is selected as the one that minimizes $\mbox{AIC}(Q)$.
\subsection{Likelihood-Ratio Test (LRT)}\label{LRT}
Errors in clustered calibration originate from the system (sky and instrumental) noise, ``clustering errors'' introduced in section \ref{Estimations of Cramer-Rao Lower Bounds}, and ``solver noise'' which is referred to as errors produced by the calibration algorithm itself. We assume that the true Jones matrices along different directions (clusters) at the same antenna are statistically uncorrelated. Note that this assumption is only made for the LRT to produce reasonable results and this assumption is not needed for clustered calibration to work. Therefore, if such correlations exist, they are caused by the aforementioned errors. Consequently, the more accurate the clustered calibration solutions are, the smaller their statistical similarities would be. Based on this general statement, the best number of clusters in a clustered calibration procedure is the one which provides us with the minimum correlations in the calibrated solutions. Note that for a fixed measurement, the correlation due to the system noise is fixed. Therefore, differences in the statistical similarities of solutions obtained by different clustering schemes are only due to ``clustering errors'' and ``solver noise''. 

To investigate the statistical interaction between the gain solutions we apply the Likelihood-Ratio Test (LRT). \\
Consider the clustered calibration solution $\widetilde{{\bf{J}}}_{pi}(\widetilde{\pmb{\theta}})$ for directions $i$, $i\in\{1,2,\ldots,Q\}$, at antennas $p$, where $p\in\{1,2,\ldots,N\}$, 
\begin{equation}
\widetilde{\textbf{J}}_{pi}=\left[\begin{array}{cc}
\widetilde{J}_{11,p}&\widetilde{J}_{12,p}\\
\widetilde{J}_{21,p}&\widetilde{J}_{22,p}\end{array}\right]_i.\label{exam}
\end{equation} 
Then, the parameter vector $\widetilde{\pmb{\theta}}_{pi}$ (corresponding to the $i$-th direction and $p$-th antenna) is obtained by
\begin{equation}
\widetilde{\pmb{\theta}}_{pi}=[\mathfrak{Re}(\widetilde{J}_{11,p})\
\mathfrak{Im}(\widetilde{J}_{11,p})\ \ldots\ \mathfrak{Re}(\widetilde{J}_{22,p})\ \mathfrak{Im}(\widetilde{J}_{22,p})]_i^T.\label{s20}
\end{equation}
Let us define for each antenna $p$ and each pair of directions $k$ and $l$, where $k$ and $l$ are belong to $\{1,2,...,Q\}$, a vector $\textbf{z}_{pkl}$ as
\begin{equation}
\textbf{z}_{pkl}=[\widetilde{\pmb{\theta}}^T_{pk}\ \widetilde{\pmb{\theta}}^T_{pl}]^T.\label{12}
\end{equation}
In fact, we are concatenating the solutions of the same antenna for two different directions (clusters) together. 

We define the null $H_0$ model as
\begin{equation}
H_0:\ \textbf{z}_{pkl}\sim \mathcal{N}(\textbf{m},\Sigma_0).\label{13.1}
\end{equation} 
where 
\begin{equation}
\textbf{m}=[\bar{\textbf{m}}(\widetilde{\pmb{\theta}}_{pk})^T\ \bar{\textbf{m}}(\widetilde{\pmb{\theta}}_{pl})^T]^T,\label{8}
\end{equation}
and 
\begin{equation}
\Sigma_\textbf{0}=\left[\begin{array}{cc}
s^2(\widetilde{\pmb{\theta}}_{pk})& 0\\
0&s^2(\widetilde{\pmb{\theta}}_{pl})\end{array}\right].\label{9}
\end{equation}
In Eq. (\ref{8}) and Eq. (\ref{9}), $\bar{\textbf{m}}$ and $s^2$ are denoting sample mean and sample variance, respectively. Note that having a large number of samples in hand, the assumption of having a Gaussian distribution for solutions is justified according to the Central Limit theorem. The structure of the variance matrix $\Sigma_\textbf{0}$ tells us that the statistical correlation between the components of the random vector $\textbf{z}_{qkl}$, or between the solutions $\widetilde{\pmb{\theta}}_{pk}$ and $\widetilde{\pmb{\theta}}_{pl}$, is zero. This is the ideal case in which there are no estimation errors. 

To investigate the validity of the null model compared with the case in which there exists some correlation between the solutions, we define the alternative $H_1$ model as
\begin{equation}
H_1:\ \textbf{z}_{pkl}\sim \mathcal{N}(\textbf{m},\Sigma_1),\label{13.2}
\end{equation}
where the variance matrix $\Sigma_1$ is given by
\begin{equation}
\Sigma_1=\left[\begin{array}{cc}
s^2(\widetilde{\pmb{\theta}}_{pk})& \mbox{Cov}(\widetilde{\pmb{\theta}}_{pk},\widetilde{\pmb{\theta}}_{pl})\\[2mm]
\mbox{Cov}(\widetilde{\pmb{\theta}}_{pk},\widetilde{\pmb{\theta}}_{pl})^T&s^2(\widetilde{\pmb{\theta}}_{pl})\end{array}\right].\label{khol}
\end{equation}\\
$\mbox{Cov}(\widetilde{\pmb{\theta}}_{pk},\widetilde{\pmb{\theta}}_{pl})$  in Eq. (\ref{khol}) is the $8\times 8$ sample covariance matrix.\\
Using the above models, the Likelihood-Ratio is defined as
\begin{equation}
\Lambda=-2\mbox{ln}\bigg(\frac{\mbox{Likelihood for null model}}{\mbox{Likelihood for alternative model}}\bigg)\label{lrt1},
\end{equation}
which has a $\chi^2$ distribution with 64 degrees of freedom. As $\Lambda$ becomes smaller, the null model, in which the statistical correlation between the solutions is zero, becomes more acceptable rather than the alternative model. Therefore, the smaller the  $\Lambda$ is, the less the clustered calibration's errors are, and vice-versa.

\section{Simulation studies}\label{example}
We use simulations to compare the performance of un-clustered and clustered calibration. Working with simulations has the advantage of having the true solutions available, which is not the case in real observations. That makes the comparison much more objective. Nevertheless, the better performance of the clustered calibration in comparison with the un-clustered ones in calibrating for real observations of LOFAR is also shown by \citet{S.K.2, Y.NCP}.

\begin{figure*}
\centering
\hspace{1cm}
\includegraphics*[width=9.5cm,height=9.5cm, viewport=400 1 1300 850]{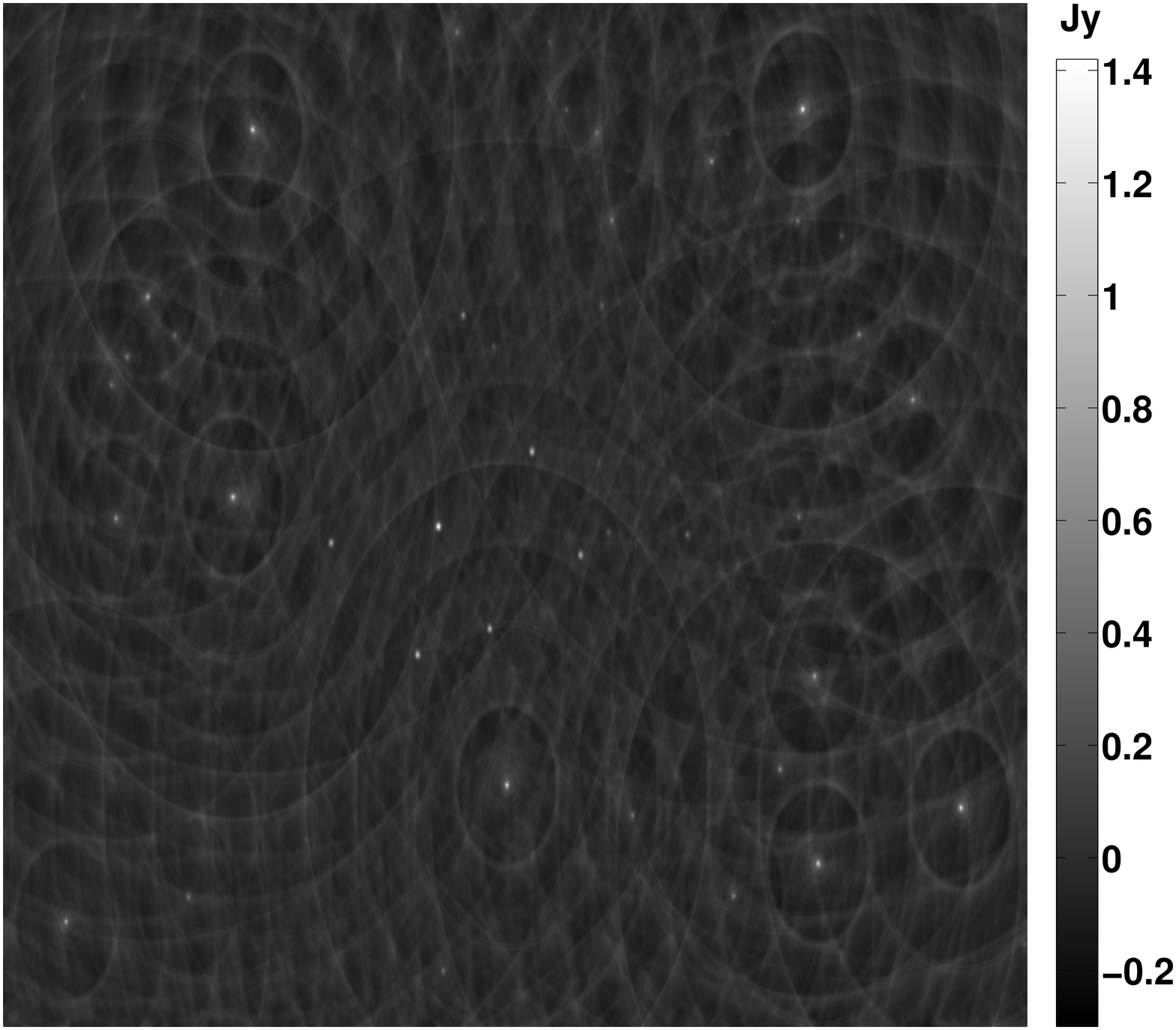}
\caption{Single channel simulated observation of fifty sources, with intensities below 3 Jy. The source positions and their brightness are following uniform and Rayleigh distributions, respectively. The image size is 8 by 8 degrees at 355 MHz. There are no gain errors and noise in the simulation.}
\label{Figure3}
\end{figure*}

We simulate an East-West radio synthesis array including 14 antennas (similar to WSRT) and an 8 by 8 degrees sky with fifty sources with low intensities, below 3 Jy. The source positions and their brightness follow uniform and Rayleigh distributions, respectively. The single channel simulated observation at 355 MHz is shown in Fig. \ref{Figure3}.

\begin{figure*}
\centering
\hspace{1cm}
\includegraphics*[width=9.5cm,height=9.5cm, viewport=400 1 1300 850]{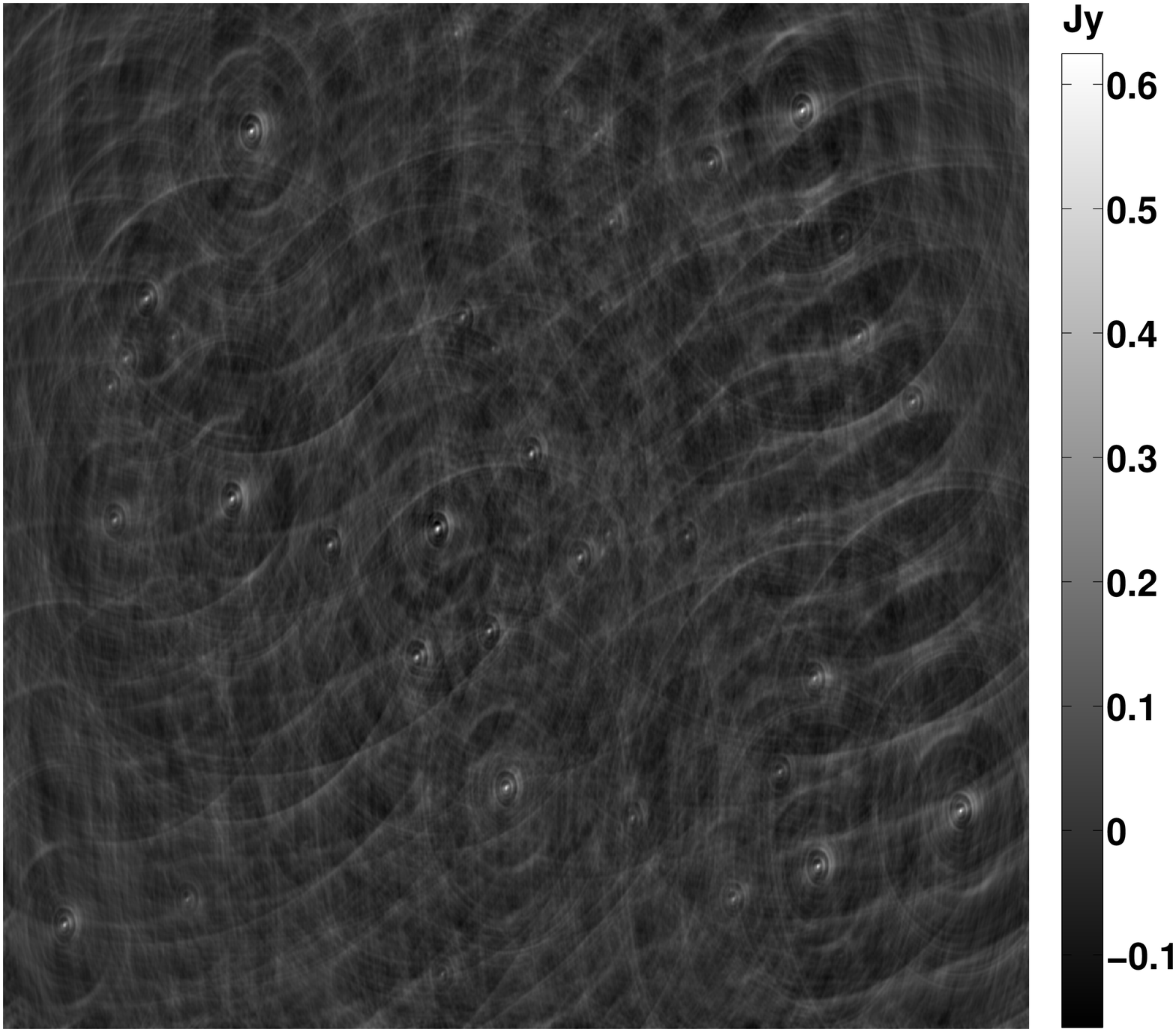}
\caption{Simulated image with added gain errors. The errors, the complex $2\times 2$ Jones matrices, are generated as linear combinations of {\it sin} and {\it cos} functions. The variation of the sky is almost uniform on small angular scales. }
\label{Figure4}
\end{figure*}

We proceed to add gain errors, multiplying source coherencies by the Jones matrices, as it is shown in Eq. (\ref{s2}), to our simulation. The amplitude and phase of the Jones matrices' elements are generated using linear combination of {\it sine} and {\it cosine} functions. We aim at simulating a sky with almost uniform variations on small angular scales. In other words, we provide very similar Jones matrices for sources with small angular separations. To accomplish this goal, for every antenna, we first choose a single direction as a reference and simulate its Jones matrix as it is explained before. Then, for the remaining forty nine sources, at that antenna, the Jones matrices (amplitude and phase terms) are that initial Jones matrix multiplied by the inverse of their corresponding angular distances from that reference direction. The result of adding such gain errors to our simulation is shown in Fig. \ref{Figure4}.

\subsection{Performance comparison of the Clustered and un-clustered calibrations at SNR=2}\label{SNR=2}
\begin{figure*}
\centering
\hspace{1cm}
\includegraphics*[width=9.5cm,height=9.5cm, viewport=400 1 1300 850]{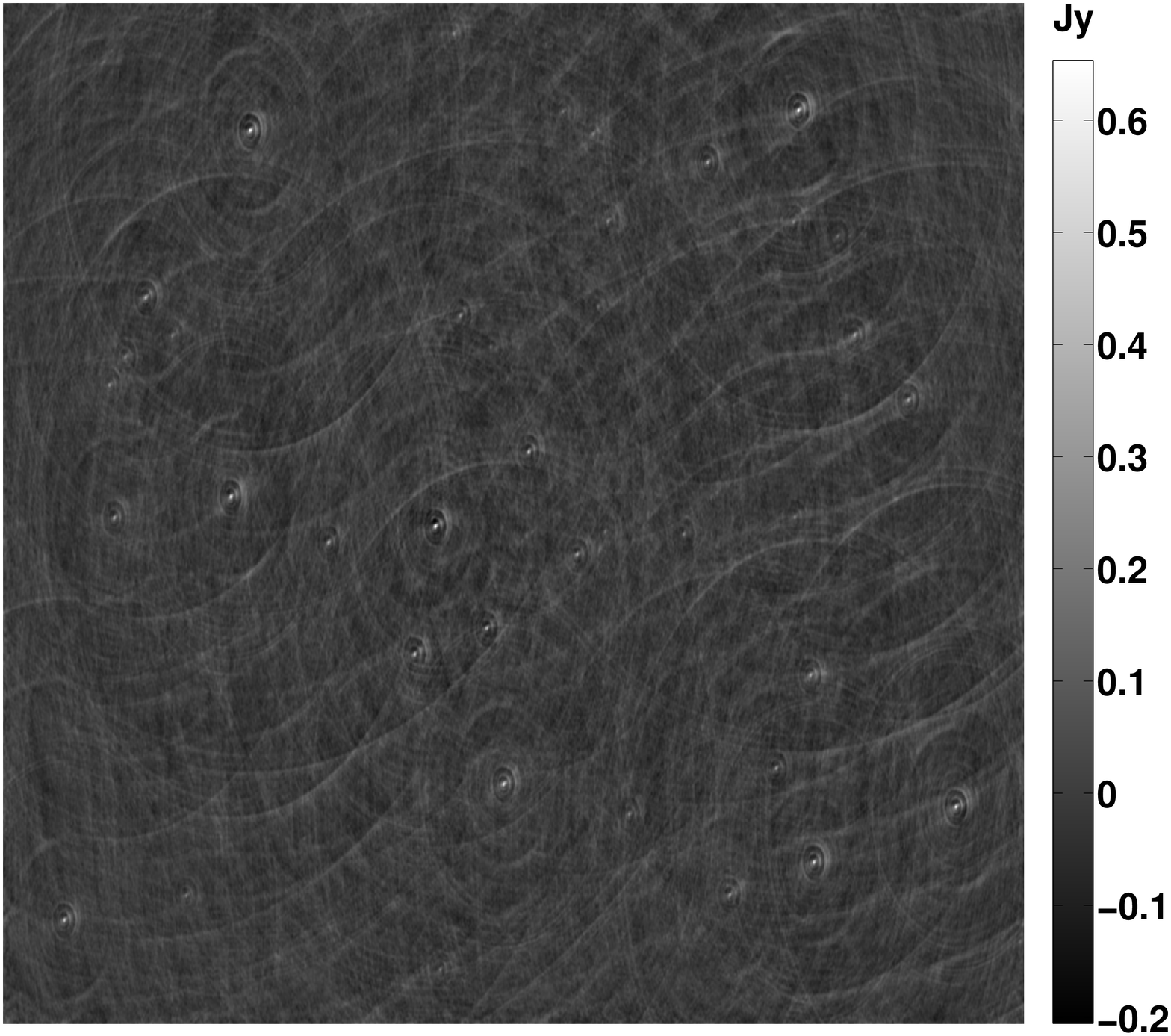}
\caption{Simulated image of sky, corrupted by gain errors and  by  noise, as in Eq. (\ref{s4}). The simulated noise vector ${\bf n}$ has zero mean complex Gaussian distribution and the SNR is equal to 2. }
\label{Figure5}
\end{figure*}

We add  noise ${\bf n}\sim\mathcal{CN}({\bf 0},\sigma^2{\bf I})$ with $\sigma^2=28$, as it is shown in Eq. (\ref{s4}), to our simulation. The result has a $SNR=2$ and is presented in Fig. \ref{Figure5}. We have chosen to present the case of $SNR=2$ since for this particular simulated observation both the divisive hierarchical and the weighted K-means clustered calibrations achieve their best performances at the same number of clusters, as will be shown later in this section.

\begin{figure}
\centering
\vspace{-0.0cm}
\hspace*{-5mm}
\includegraphics[width=9.5cm,height=6cm]{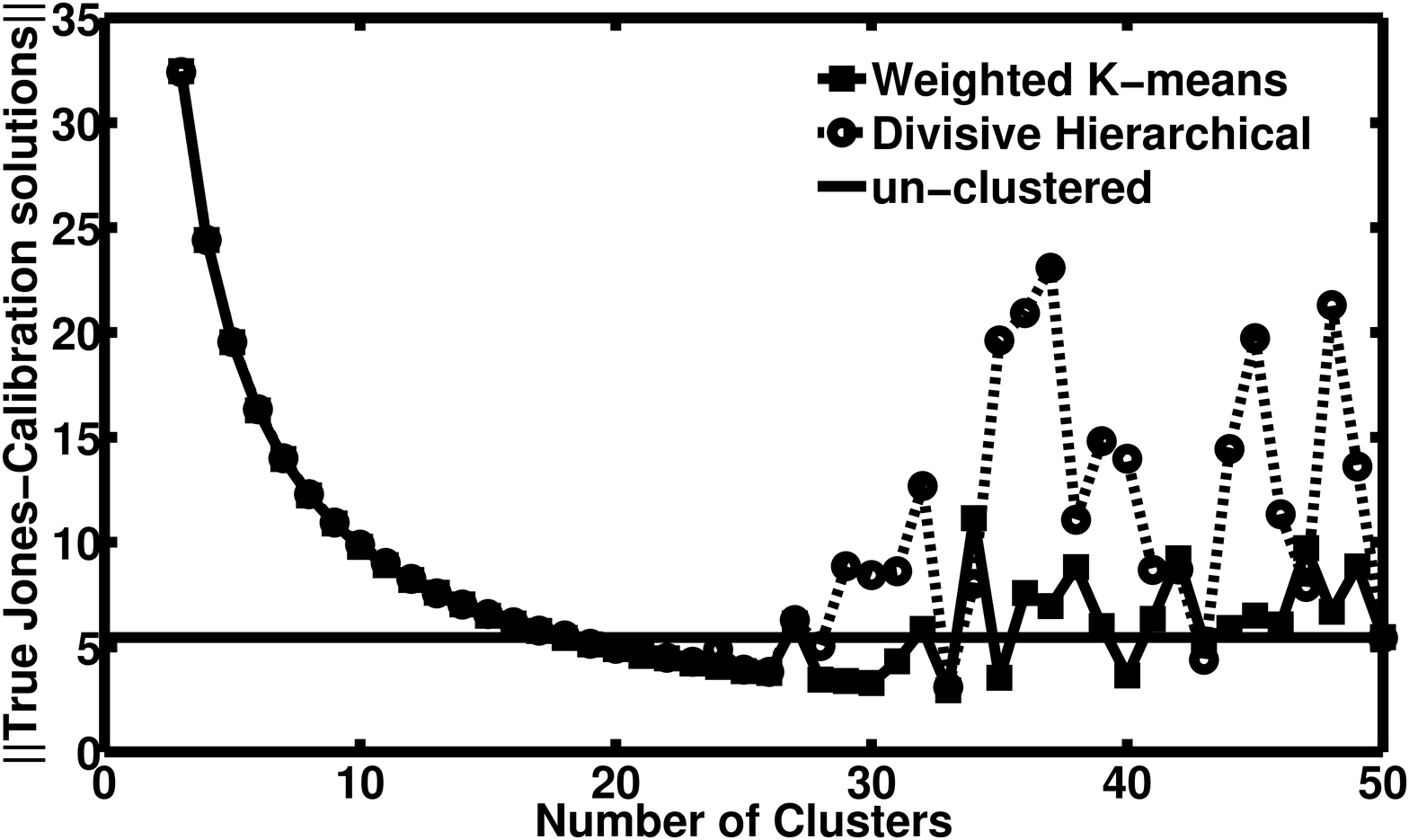}
\vspace{-0.5cm}
\caption{The average Frobenius distance between the simulated (true) Jones matrices and the solutions of clustered and un-clustered calibrations. The two curves represent clustered calibration via divisive hierarchical and weighted K-means clustering algorithms. By increasing the number of clusters, for both clustering methods, this distance is decreased and gets its minimum approximately at thirty three clusters. After that, it is increased till the fifty individual sources. That shows that the best performance of the clustered calibration is at around thirty three clusters. }
\label{Figure6}
\end{figure}

We apply un-clustered and clustered calibration on the simulation to compare their efficiencies. The fifty sources are grouped into $Q\in\{3,4,\ldots,49\}$ number of clusters, using the proposed divisive hierarchical and weighted K-means clustering algorithms. Self-calibration is implemented via Space Alternating Generalized Expectation Maximization (SAGE) algorithm \citep{J.A.1, S.2, S.K} with nine iterations. Plots of the averaged Frobenius distance between the simulated (true) Jones matrices and the obtained solutions is shown in Fig. \ref{Figure6}. As we can see  in Fig. \ref{Figure6}, for both clustering schemes, increasing the number of clusters decreases this distance and the minimum is reached at approximately thirty three clusters ($Q=33$). Beyond this number of clusters, it increases until the fifty individual sources become individual clusters. This shows that the best performance of both the divisive hierarchical and the weighted K-means clustered calibrations is approximately at thirty three clusters and is superior to that of the un-clustered calibration. \\
The Frobenius distance curves in Fig. \ref{Figure6}, the MCRLB curve in Fig. \ref{figure12}, and the SINR curve in Fig. \ref{figure2} illustrate that clustered calibration with an extremely low number of clusters does not necessarily perform better than the un-clustered calibration. The reason is that when there are only a small number of clusters, the interference, or the so-called ``clustering errors'' introduced in section \ref{Estimations of Cramer-Rao Lower Bounds}, is relatively large. Therefore, the effect of this interference dominates the clustering of signals. On the other hand, we reach the theoretical performance limit approximately after twenty five number of clusters and therefore increasing the number beyond this point gives highly variable results, mainly because we are limited by the number of constraints as opposed to the number of parameters that we need to solve for. But, this is not the case for the plots in Fig. \ref{figure12} and Fig. \ref{figure2}. The reason is that the MCRLB results of Fig. \ref{figure12} as well as the SINR results of Fig. \ref{figure2} are obtained by Monte-Carlo method with iterations over fifty different sky and noise realizations. However, Fig. \ref{Figure6} is limited to the presented specific simulation with only one sky and one noise realization.

\begin{figure*}
\centering
\hspace{1cm}
\includegraphics*[width=9.5cm,height=9.5cm, viewport=400 1 1300 850]{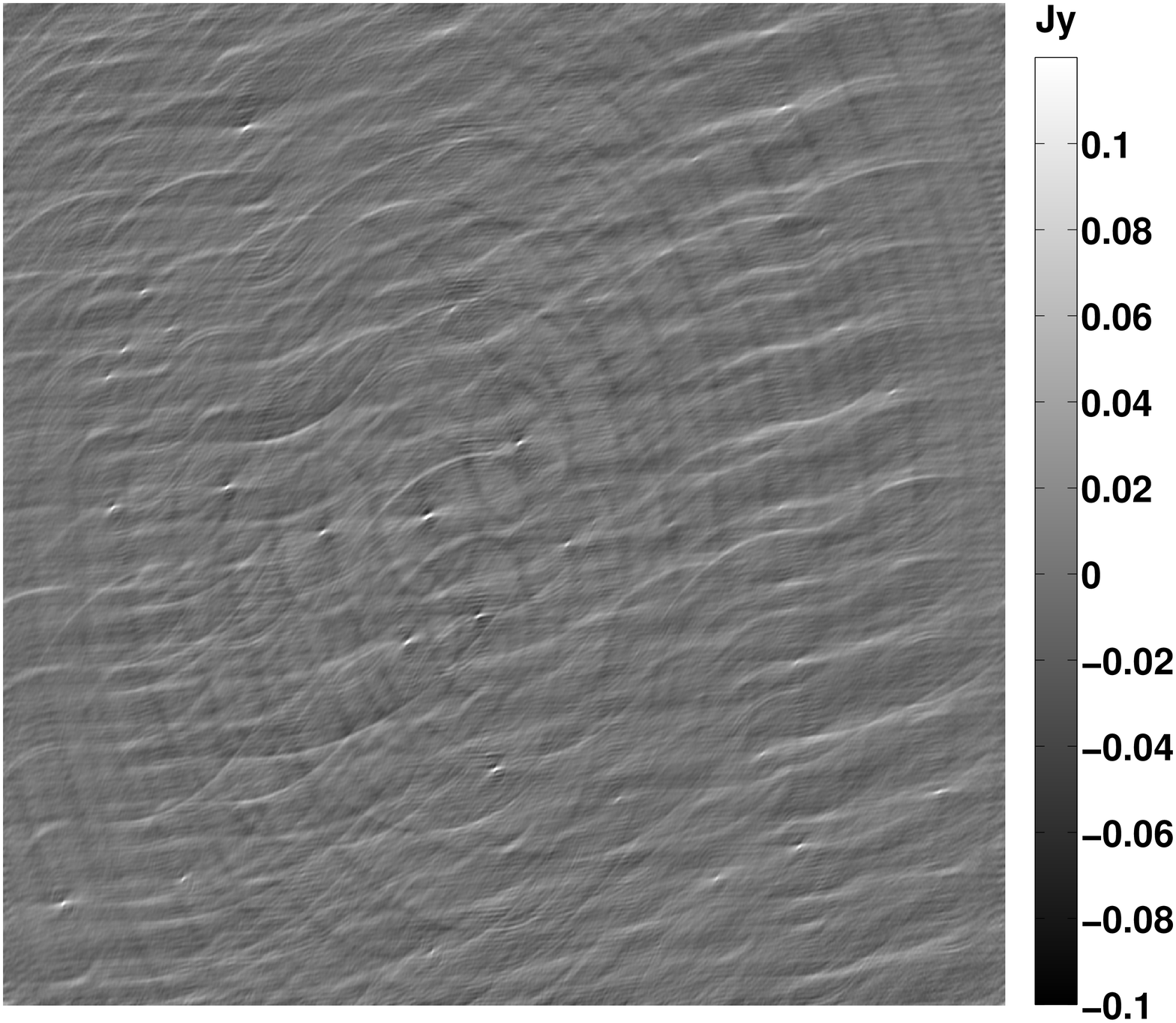}
\caption{Residual image of the un-clustered SAGE calibration for fifty sources. The sources are almost not subtracted at all and there are significant residual errors around them. The residuals have asymmetric Gaussian distribution with variance $\sigma^2=82.29$ which is much larger than the true noise variance $\sigma^2=10.85$.}
\label{Figure111x}
\end{figure*}

\begin{figure*}
\centering
$\begin{array}{cc}
\hspace*{3mm}\includegraphics*[width=9cm,height=9.3cm, viewport=400 1 1300 850]{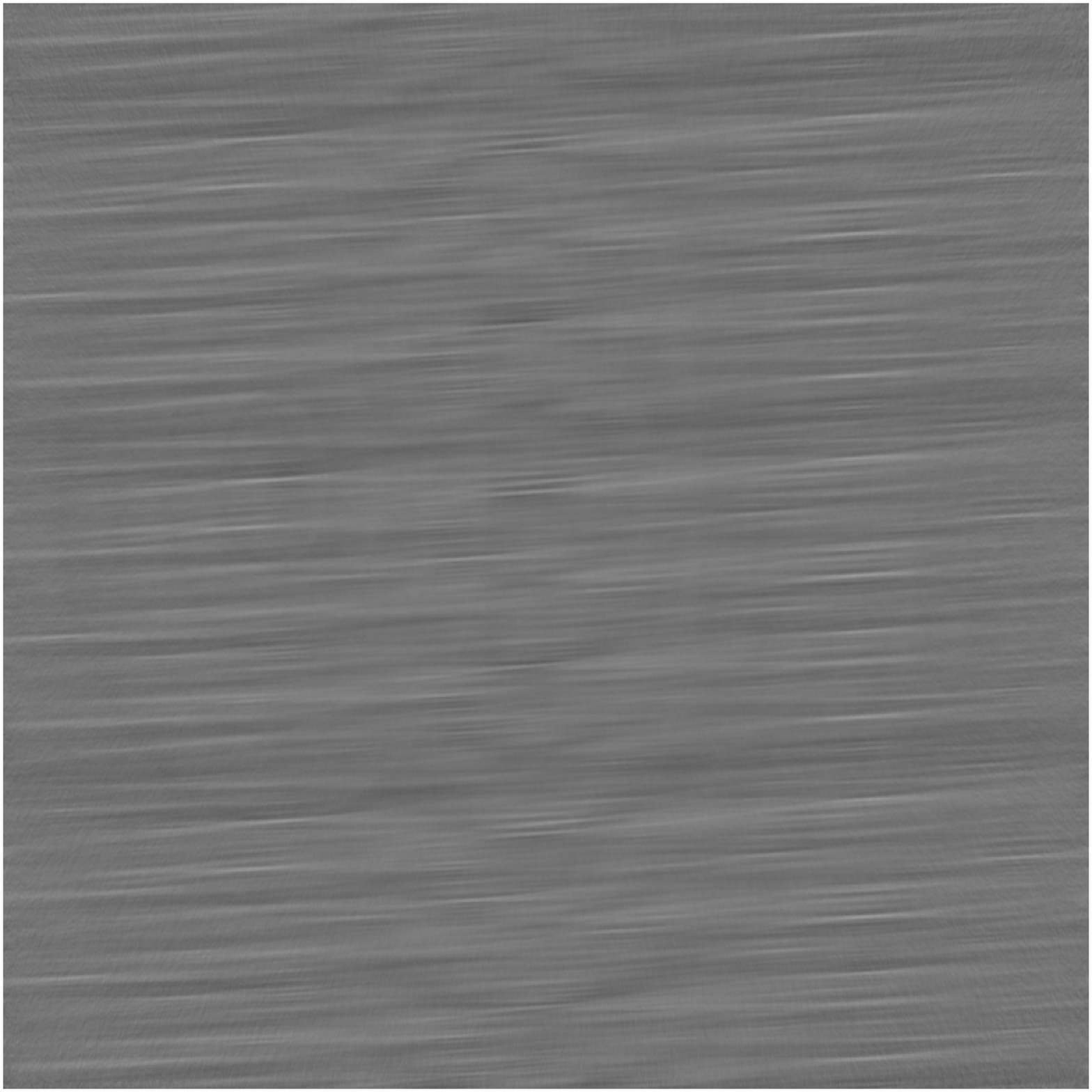}&
\hspace*{-1.2cm}\includegraphics*[width=9cm,height=9.5cm, viewport=400 1 1300 850]{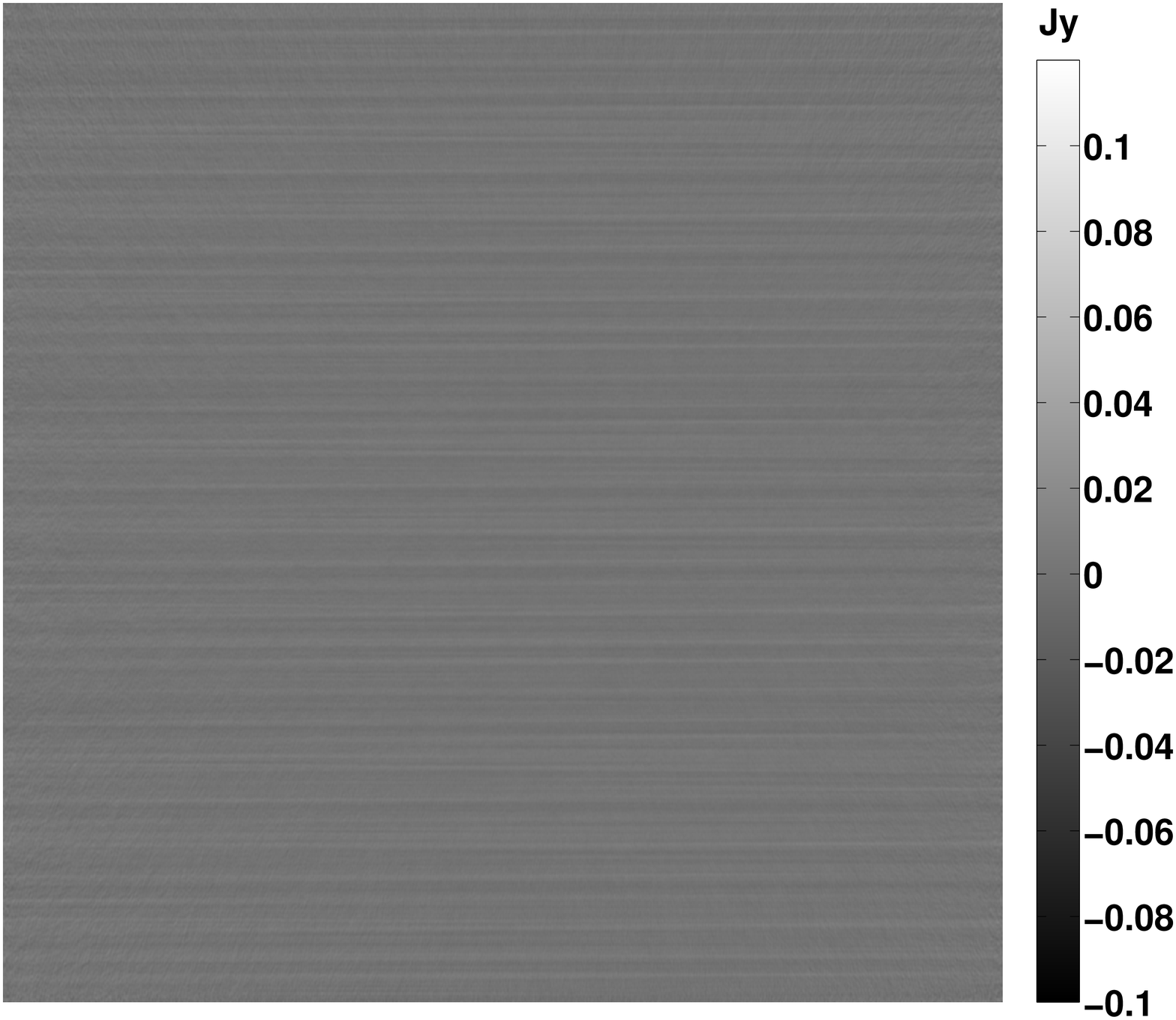}
\end{array}$
\caption{ The residual images of the clustered calibration using hierarchical (right) and Weighted K-means (left) clustering methods with thirty three source clusters. Calibration is implemented by SAGE algorithm. The sources are subtracted perfectly and the residuals converge to the simulated background noise distribution. The hierarchical clustering and Weighted K-means residuals follow symmetric zero mean Gaussian distributions with $\sigma^2=20.17$ and $  \sigma^2=18.76$, respectively, where the simulated noise distribution is, $\mathcal{CN}({\bf 0},10.85{\bf I})$.}\label{Figure11}
\end{figure*}

The residual images of the un-clustered calibration as well as the divisive hierarchical and weighted K-means clustered calibrations for $Q=33$ are shown by Fig. \ref{Figure111x} and Fig. \ref{Figure11}, respectively. As it is shown by Fig. \ref{Figure111x}, in the result of un-clustered calibration, the sources are almost not subtracted at all and there is a significant residual error remaining. The residuals have asymmetric Gaussian distribution with variance $\sigma^2=82.29$ which is much larger than the simulated (true) noise variance $\sigma^2=10.85$. On the other hand, the sources have been perfectly subtracted in the case of clustered calibration, Fig. \ref{Figure11}, and the residuals converge to the simulated background noise distribution. The residuals of the divisive hierarchical and Weighted K-means clustered calibrations follow symmetric zero mean Gaussian distributions with $\sigma^2=20.17$ and $\sigma^2=18.76$, respectively. These variances are closer to the simulated one $\sigma^2=10.85$ and this indicates the promising performance of clustered calibration. As we can see, hierarchical clustered calibration provides a slightly better result compared to the K-means one. This is due to the fact that hierarchical clustering constructs clusters of smaller angular diameters and thus it assigns the same calibration solutions to the sources with smaller angular separations. 

\begin{figure}
\centering
\vspace{-0.0cm}
\hspace*{-5mm}
\includegraphics[width=9.5cm,height=6cm]{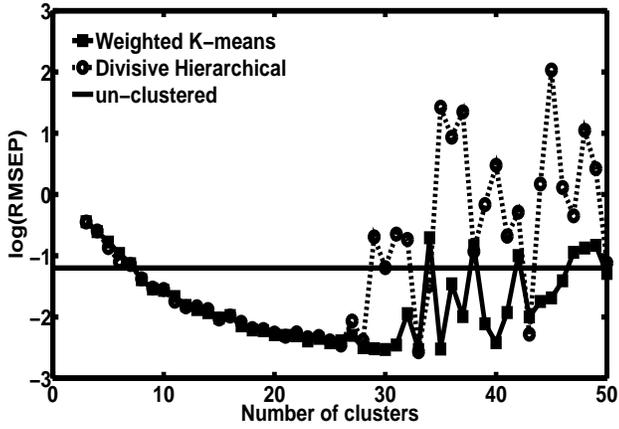}
\vspace{-0.5cm}
\caption{The RMSEP for clustered and un-clustered calibrations. The results are obtained using a base ten logarithmic scale. The two curves are corresponding to clustered calibration via divisive hierarchical and weighted K-means clustering algorithms. By increasing the number of clusters, the results are decreased and the minimum result is obtained at around thirty three clusters. After that, the results are increased till the fifty individual sources. That shows the superior performance of the clustered calibration compared to the un-clustered one. The best performance of the clustered calibration for both of the applied clustering methods is  at around thirty three number of clusters. }
\label{Figure7}
\end{figure}

We also calculate the Root Mean Squared Error of Prediction (RMSEP) to assess the performance of clustered and un-clustered calibrations' non-linear regressions. The results of log(RMSEP), presented by Fig. \ref{Figure7}, also justify that the best efficiencies of the hierarchical and K-means clustered calibrations are obtained at thirty three number of clusters. But, note that there is a difference between the behavior of log(RMSEP) plot of Fig. \ref{Figure7} and the plots of MCRLB, SINR, and Frobenius distance between the simulated Jones matrices and solutions in Fig. \ref{figure12}, Fig. \ref{figure2}, and Fig. \ref{Figure6}, respectively. In Fig. \ref{Figure7}, log(RMSEP) of clustered calibration is less than that of un-clustered calibration, even for extremely low number of clusters. This means that even with those low number of clusters, clustered calibration performs better than the un-clustered calibration. This is somewhat in disagreement with the scenarios shown in Fig. \ref{figure12}, Fig. \ref{figure2}, and Fig. \ref{Figure6}. For a  better understanding of the reason behind this contrast, first lets see how the residual errors are originated. \\
Based on Eq. (\ref{sa4}) and Eq. (\ref{nn4}), in the clustered calibration strategy, we have
\begin{equation}
{\bf y}=\sum_{i=1}^Q \widetilde{\bf s}_i(\widetilde{\pmb{\theta}})+\sum_{i=1}^K {\pmb \Gamma}_i+{\bf n},\label{new1}
\end{equation}
After executing a calibration for the above data model, there is a distance between the target parameters $\widetilde{\pmb{\theta}}$ and the estimated solutions $\widehat{\widetilde{\pmb{\theta}}}$. This is the so-called ``solver noise'', mentioned in section \ref{LRT}. Thus, the residuals are given as 
\begin{equation}
{\bf y}-\sum_{i=1}^Q \widetilde{\bf s}_i(\widehat{\widetilde{\pmb{\theta}}})=\sum_{i=1}^Q \{\widetilde{\bf s}_i(\widetilde{\pmb{\theta}})-\widetilde{\bf s}_i(\widehat{\widetilde{\pmb{\theta}}})\}+\sum_{i=1}^K {\pmb \Gamma}_i+{\bf n}.\label{new2}
\end{equation}
From Eq. (\ref{new2}), we immediately see that the background noise ${\bf n}$ is fixed and the ``clustering errors'' are calculated for all the sources as $\sum_{i=1}^K {\pmb \Gamma}_i$. However, since we solve only for $Q$ directions and not for all the $K$ sources individually, the ``solver noise'' part, $\sum_{i=1}^Q \{\widetilde{\bf s}_i(\widetilde{\pmb{\theta}})-\widetilde{\bf s}_i(\widehat{\widetilde{\pmb{\theta}}})\}$, is also calculated only for $Q$ clusters and not for all the $K$ sources. It is clear that for a very small $Q$, this term could be much less than for $Q\simeq K$. Therefore, in Fig. \ref{Figure6}, the result of RMSEP at a very low number of clusters is still less than the ones at $Q=K$. When $Q$ is not at the two extremes of being very small or very large (almost equal to the number of individual sources), then the result of RMSEP is promising. Moreover, applying more accurate calibration methods or increasing the number of iterations, the ``solver noise''  will decrease and subsequently, we expect the RMSEP curve to have the same behavior as the curves of Fig. \ref{figure2} and Fig. \ref{Figure6}.

\subsection{Optimum number of clusters for SNR=2}
\begin{figure}
\centering
\vspace{-0.0cm}
\hspace*{-5mm}
\includegraphics[width=9.5cm,height=6cm]{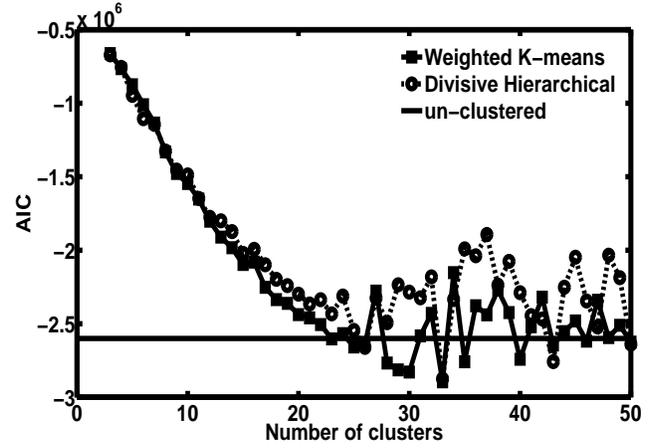}
\vspace{-0.5cm}
\caption{AIC plot for clustered and un-clustered calibrations. Both the weighted K-means and divisive hierarchical clustered calibrations get their minimum AIC at about thirty three clusters. This illustrates that their best performances are obtained at this number of clusters. Also, their AIC results at thirty three clusters is lower than the un-clustered calibration's AIC, which shows their better performances compared to the un-clustered calibration. }
\label{Figure9}
\end{figure}

\begin{figure}
\centering
\vspace{-0.0cm}
\hspace*{-5mm}
\includegraphics[width=9.5cm,height=6cm]{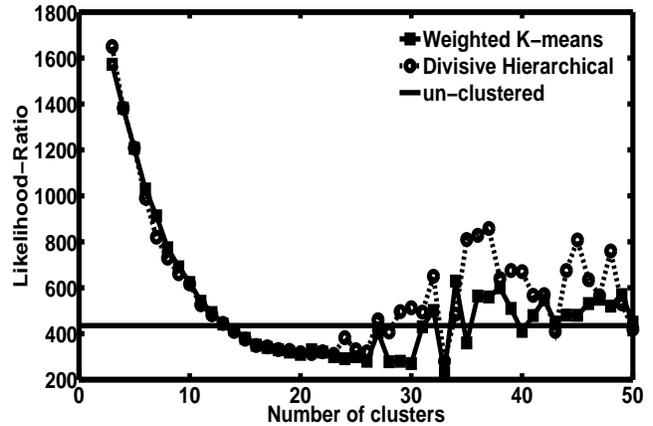}
\vspace{-0.5cm}
\caption{Likelihood-ratio of the gain solutions obtained by clustered and un-clustered calibrations. In the both cases of weighted K-means and divisive hierarchical clustered calibrations, the minimum Likelihood-ratio values belong to approximately thirty three number of clusters. These minimums are also lower than the un-clustered calibration's Likelihood-ratio result. Therefore, clustered calibration via both the clustering methods performs better than the un-clustered calibration and it achieves the best accuracy in its solutions at thirty three clusters. }
\label{Figure10}
\end{figure}

We utilize AIC and LRT to select the optimum number of clusters for which clustered calibration achieves its best performance. The methods are applied to our simulation for the case of SNR=2. The AIC and LRT results are shown in Fig. \ref{Figure9} and Fig. \ref{Figure10}, respectively. They both agree on $Q=33$ as the optimum number of clusters for the divisive hierarchical and Weighted K-means clustered calibrations. Likelihood-Ratio plot of Fig. \ref{Figure10} has almost the same behavior as the plot of Frobenius distance between the simulated Jones matrices and the obtained solutions presented by Fig. \ref{Figure6}. The reason is that the results of the both plots are obtained using the solutions themselves as the input data. However, since AIC results are computed using the residual errors as inputs, which is also the case for obtaining the RMSEP curves of Fig. \ref{Figure7}, AIC curves of Fig. \ref{Figure9} are slightly steeper than the Frobenius distance between the simulated Jones matrices and solutions and the Likelihood-Ratio curves of Fig. \ref{Figure6} and Fig. \ref{Figure10}, respectively. 

\subsection{Clustered calibration's efficiency at different SNRs}
\begin{figure}
\centering
\vspace{-0.0cm}
\hspace*{-5mm}
\includegraphics[width=9.5cm,height=6cm]{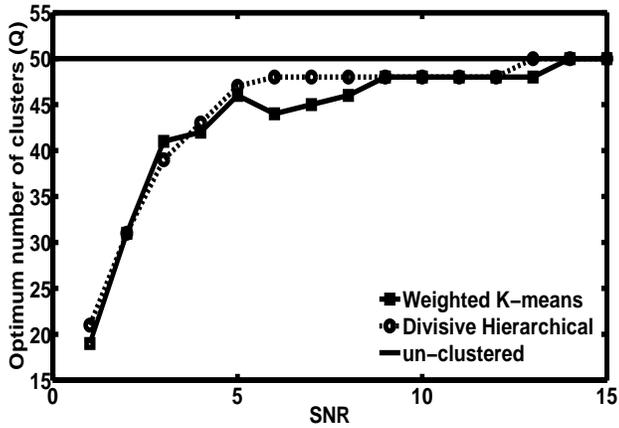}
\vspace{-0.5cm}
\caption{The optimum number of clusters, on which the best performance of clustered calibration is obtained, at different SNRs. For low SNRs, the efficiency of the clustered calibration is superior to the un-clustered calibration. As the SNR gets higher, clustered calibrations achieve their best solutions utilizing a higher number of clusters. Finally, when the SNR is high enough, the performance of un-clustered calibration becomes better than the clustered one.}
\label{Figure8}
\end{figure}

We start changing the noise in our simulation to see how it effects the clustered calibration's efficiency. We simulate the cases for which SNR$\in\{1,2,\ldots,15\}$ and apply clustered calibration on them. Since the sky model does not change, the clusters obtained by divisive hierarchical and weighted K-means methods for the case of SNR$=2$ remain the same. Fig. \ref{Figure8} shows the optimum number of clusters, on which the best performances of clustered calibrations are obtained for those different SNRs. As we can see in Fig. \ref{Figure8}, for low SNRs, the optimum $Q$ is small. By increasing the SNR, the optimum $Q$ is increased till it becomes equal to the number of all the individual sources that we have in the sky, i.e., $K$. This means that when the SNR is very low, the benefit of improving signals by clustering sources is much higher than the payoff of introducing ``clustering errors'' in a clustered calibration procedure. Therefore, clustered calibration for $Q\ll K$ has a superior performance compared to the un-clustered calibration. While for a high enough SNR, the situation becomes the opposite. In this case, the un-clustered calibration achieves better results compared to clustered calibration having the disadvantage of introducing ``clustering errors''.

\subsubsection{Empirical estimation of SINR}
Having the results of Fig. \ref{Figure8} in hand, we could find an empirical model for estimating the optimum number of clusters for various SNRs. Note that by changing the observation and the instrument characteristics, this model will also be changed. 

As it is explained in section \ref{Analysis}, the best performance of clustered calibration is obtained when the SINR is at its highest level. Fig. \ref{Figure8} shows the number of clusters on which the maximum SINR was obtained, where the signal (sky) and the noise powers are known a priori. Thus, the only unknown for estimating the SINR is the interference, or the ``clustering errors'', for which we need to have a prediction model. After estimating the SINR using this model, finding the optimum $Q$ will be straightforward.

\begin{figure}
\centering
$\begin{array}{c}
\hspace*{-8mm}\includegraphics[width=9.5cm,height=6cm]{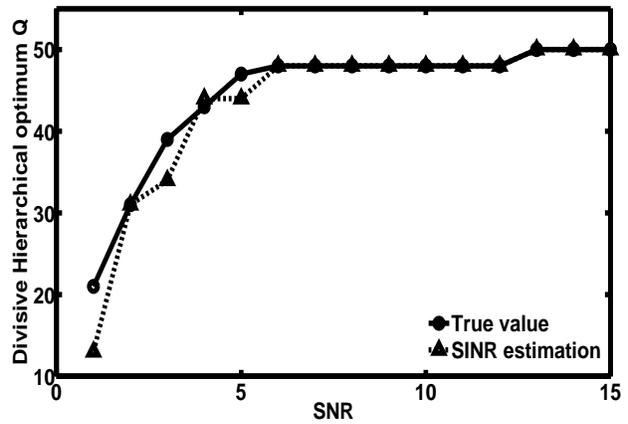}\\
\hspace*{-8mm}\includegraphics[width=9.5cm,height=6cm]{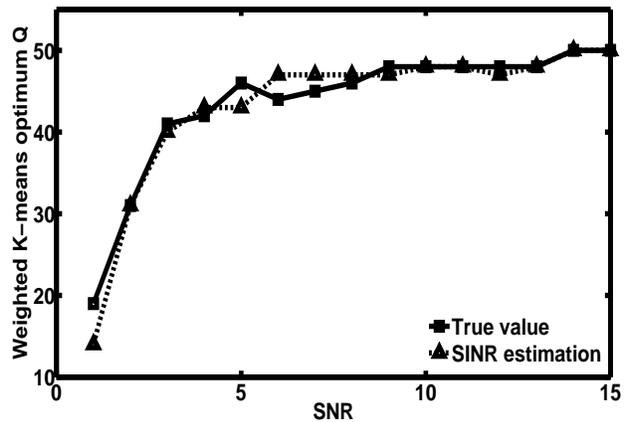}
\end{array}$
\caption{ Optimum number of clusters at which the divisive hierarchical (top) and Weighted K-means (bottom) clustered calibrations perform the best. For both of the clustering methods, the results obtained by SINR estimations mostly match the true optimum number of clusters. }\label{Figure12}
\end{figure}

Consider the definition of ``clustering errors'' given by Eq. (\ref{s18}). It is logical that for every source, the difference between its true Jones matrix and the clustered calibration solution, $||{\bf J}-\widetilde{\bf J}||$, is a function of the angular distance between the source and the centroid of the cluster that it belongs to. Based on this and using Eq. (\ref{s18}) and Eq. (\ref{s8}), for the interference of the $i$-th cluster at baseline $p-q$ we assume that
\begin{equation}
\sum_{l\in L_i}{\bf \Gamma}_{l\{pq\}}\sim\mathcal{CN}({\bf 0},\eta ||\widetilde{\bf C}_{i\{pq\}}||^2 \{D(L_i)\}^\nu {\bf I}_2),\label{kheng}
\end{equation}
where $\eta$ and $\nu$ are unknowns. Eq. (\ref{kheng}), in fact, considers an interference power (variance) of $\eta||\widetilde{\bf C}_{i\{pq\}}||^2 \{D(L_i)\}^\nu$ for every $i$-th cluster, $i\in\{1,2,\ldots,Q\}$, at baseline $p-q$. Assuming the interferences of different clusters to be statistically independent from each other, and bearing in mind that the baseline's additive noise ${\bf N}_{pq}$ has also a complex Gaussian distribution independent from those interferences', then the noise plus interference power for the $i$-th cluster at baseline $p-q$ is obtained by
\begin{equation}
\eta||\widetilde{\bf C}_{i\{pq\}}||^2 \{D(L_i)\}^\nu+||{\bf N}_{pq}||^2.\label{s778}
\end{equation}
Fitting suitable $\eta$ and $\nu$ to Eq. (\ref{s778}), the SINR for the $i$-th cluster at baseline $p-q$ is equal to the cluster's signal power, $||\widetilde{\bf C}_{i\{pq\}}||^2$, divided by the result of Eq. (\ref{s778}). Subsequently, estimation of E\{SINR$_c$\} will be straight forward where the expectation is calculated with respect to all the source clusters and all the baselines. Note that simulation provides us with the true noise power, $||{\bf N}_{pq}||^2$. In the case of having a real observation, this power could be estimated by Eq. (\ref{aic2}). \\
Fig. \ref{Figure12} shows the number of clusters on which divisive hierarchical and weighted K-means clustered calibrations achieve their maximum estimated E\{SINR$_c$\}. The results are calculated for SNR$\in\{1,2,\ldots,15\}$. For the hierarchical clustering $\eta=1550$ and $\nu=0.3$, and for the K-means clustering $\eta=2500$ and $\nu=0.003$. As we can see, for both the clustering methods, these maximum E\{SINR$_c$\}s are mostly obtained at the true optimum number of clusters for which clustered calibration performed the best. Introducing more refined models compared to Eq. (\ref{kheng}) could even improve the current result.

\subsection{Realistic sky models}
So far, we have limited our studies to sky models in which the brightness and position of the radio sources follow  Rayleigh and uniform distributions, respectively. These characteristics provide us with a smooth and uniform variation of flux intensities in our simulated skies. In such a case, the effects of the background noise on the faintest and the strongest signals are almost the same. Therefore, if clustered calibration performs better than the un-clustered calibration that would be only based on upgrading the signals against the noise. Although, in nature, we mostly deal with the sky models in which the distribution of the  flux intensities is a power law, with a steep slope, and the spatial distribution is Poisson. Hence, there exist a few number of very bright sources, whose signals are considerably stronger than the others, and they are sparse in the field of view. The corruptions of the background noise plus the interferences of the strong signals of those few bright sources make the calibration of the other faint point sources impractical. Thus, there is the need for utilizing the clustered calibration which applies the solutions of the bright sources to their closed by fainter ones or solves for upgraded signals obtained by adding up a group of faint signals together. This has been shown by \citet{S.K.2, Y.NCP}, when comparing the efficiency of the clustered and un-clustered calibrations on LOFAR real observations. In this section, using simulations, we also reveal the superiority of clustered calibration compared to the un-clustered calibration for such sky models.  

\begin{figure*}
\centering
\hspace{1cm}
\includegraphics*[width=9.5cm,height=9.5cm, viewport=400 1 1300 850]{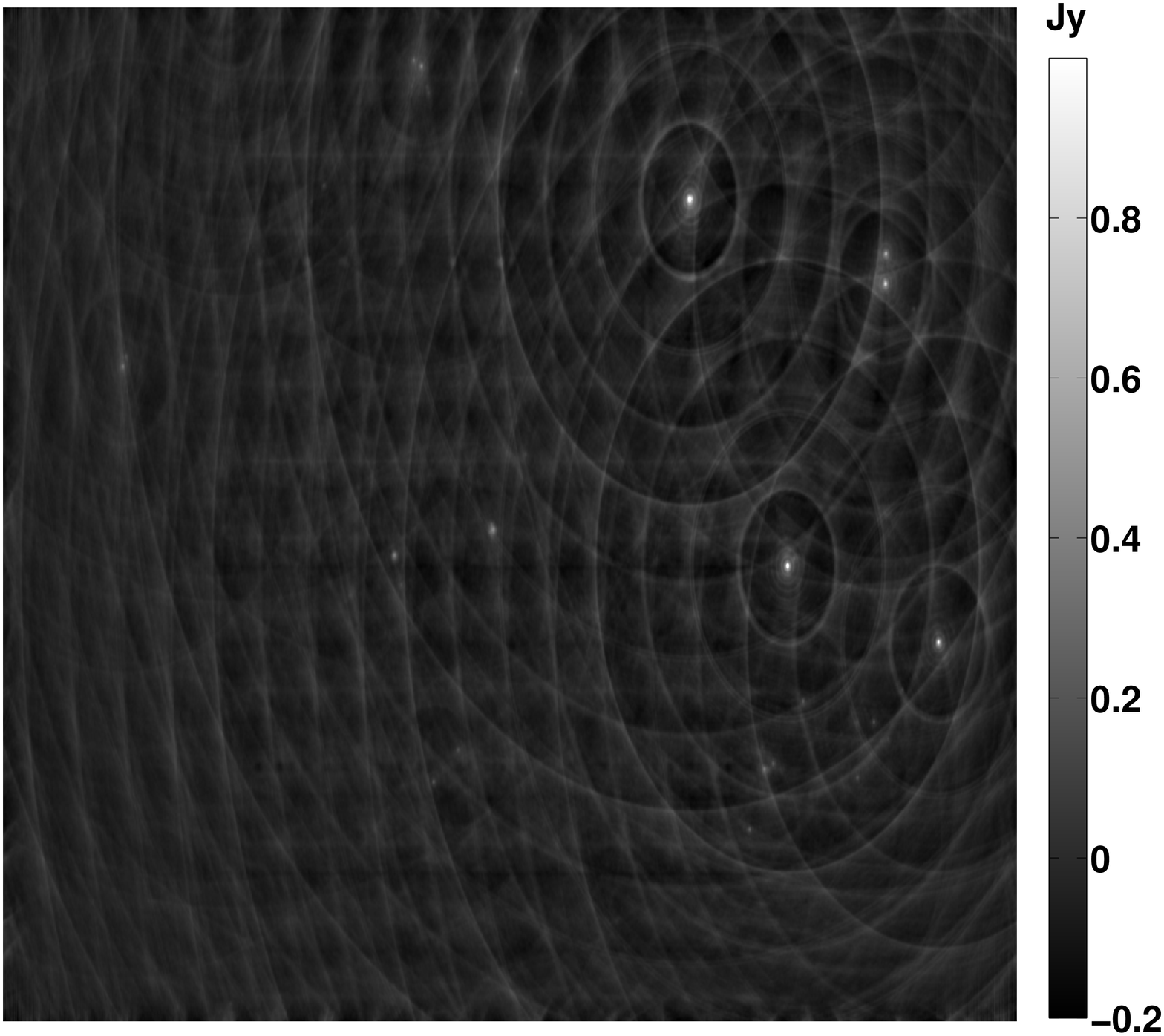}
\caption{Simulated observation of fifty two sources which are obtained by modified \citet{jelic08} foreground model. The corrupting gain errors are generated as linear combinations of \emph{sin} and \emph{cos} functions. The image size is 8 by 8 degrees and the additive thermal noise is a zero mean Gaussian noise with a standard deviation of 3 mJy. }
\label{Figure33}
\end{figure*}

We simulate a sky of 52 radio point sources which are obtained by modified \citet{jelic08} foreground model. The brightness distribution of the point sources follows the source count function obtained at 151 MHz \citep{willott01}, while the angular clustering of the sources are characterized by a typical two-point correlation function, 
\begin{equation}
\rho(d)=Ad^{-0.8}.\label{vib}
\end{equation}
In (\ref{vib}), $\rho$ is the two point correlation function, $d$ is the angular separation, and $A$ is the normalization amplitude of $\rho$. The flux cut off is 0.1 Jy. \\
We corrupt the signals with gain errors which are linear combinations of \emph{sin} and \emph{cos} functions, as in our previous simulations. At the end, a zero mean Gaussian thermal noise with a standard deviation of 3 mJy is added to the simulated data. The result is shown in Fig. \ref{Figure33}.  In Fig. \ref{Figure33}, all the bright sources are concentrated on the right side of the image, rather than being uniformly distributed in the field of view, and the rest of the sources are so faint that are almost invisible. 

\begin{figure}
\centering
\vspace{-0.0cm}
\hspace*{-5mm}
\includegraphics[width=9.5cm,height=6cm]{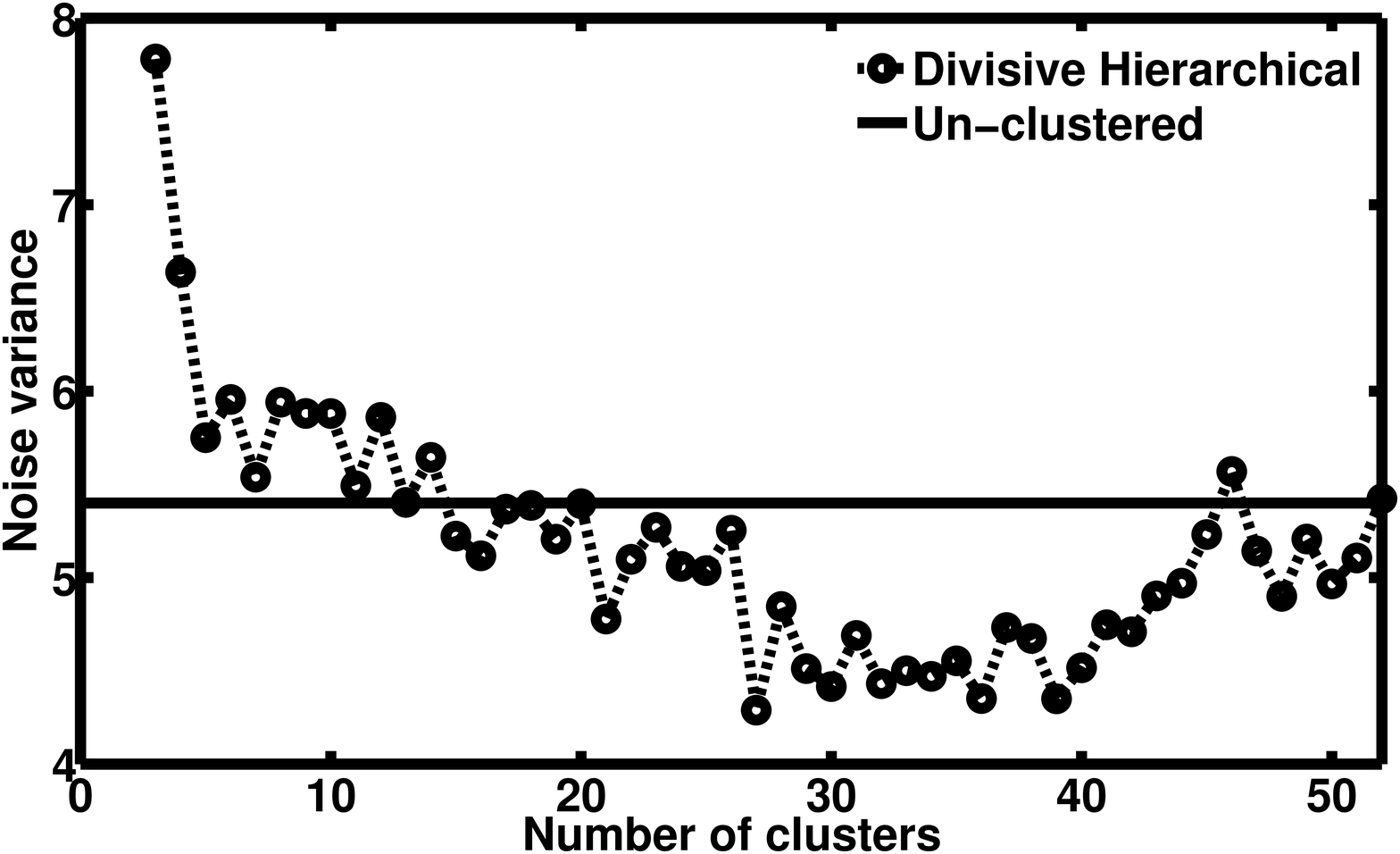}
\vspace{-0.5cm}
\caption{The noise variances of the residual images, obtained by clustered and un-clustered calibrations, in mJy. The level of the residual noise obtained by the clustered calibration for $Q\in\{15,16,\ldots 45\}$ number of clusters stands below the result of the un-clustered calibration. That reveals the superior performance of the clustered calibration in comparison with the un-clustered one. The best result of the clustered calibration which achieves the minimum noise level is at  $Q=27$ number of clusters. }
\label{Figure88}
\end{figure}

We apply the clustered and un-clustered calibrations on $Q\in\{3,4,\ldots 51\}$ number of clusters and $K=52$ number of individual sources, respectively. The clustering method used is the divisive hierarchical and the calibrations are executed via SAGE algorithm with nine number of iterations. The residual noise variances obtained are demonstrated in Fig. \ref{Figure88}. As Fig. \ref{Figure88} shows, the level of the residual noise obtained by the clustered calibration for $Q\in\{15,16,\ldots 45\}$ number of clusters is always below the result of the un-clustered calibration. This proves the better performance of the clustered calibration. The best result of the clustered calibration, with the minimum noise level, is achieved for $Q=27$ number of clusters. 

\begin{figure*}
\centering
$\begin{array}{cc}
\includegraphics*[width=9cm,height=9.3cm, viewport=380 1 1271 850]{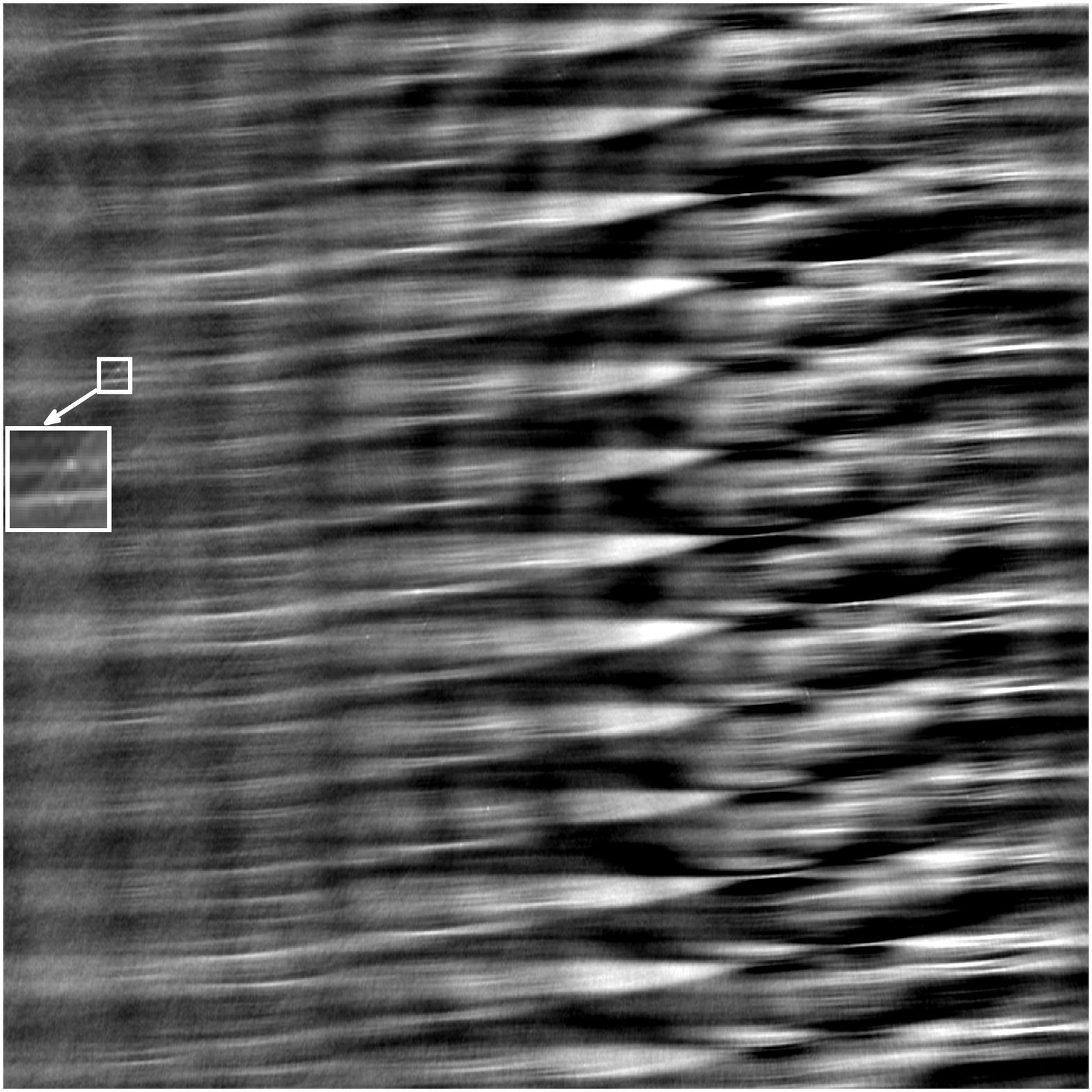}&
\hspace*{-1.2cm}\includegraphics*[width=9.3cm,height=9.5cm, viewport=380 1 1289 850]{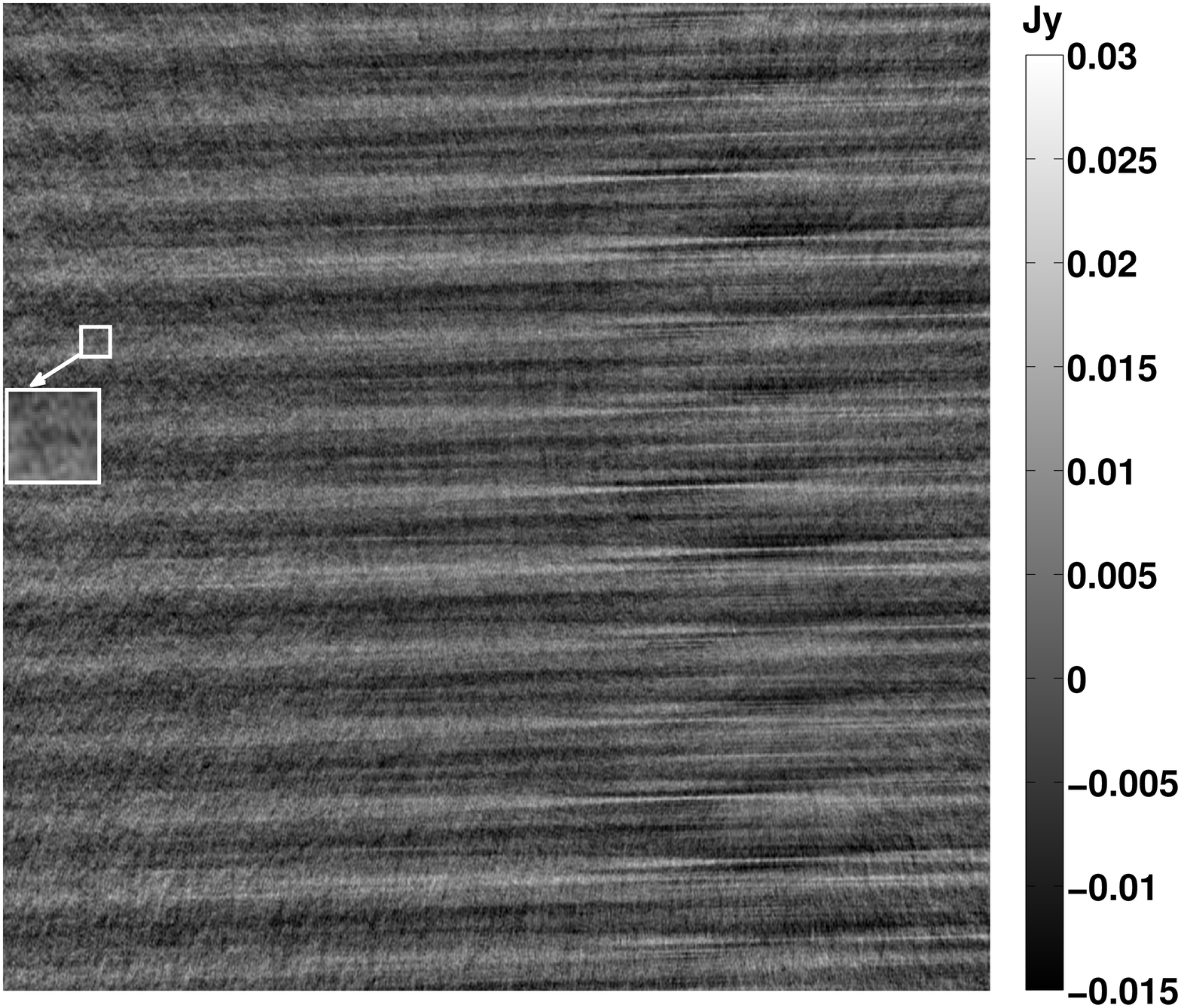}
\end{array}$
\caption{ The residual images of the clustered calibration for $Q=27$ number of clusters (right) and the un-clustered calibration for $K=52$ (left). Calibration is implemented by SAGE algorithm with nine number of iterations and the clustering method applied is divisive hierarchical. In the right side of the residual image of the un-clustered calibration we see stripes of over and under estimations due to problematic performance of the un-clustered calibration in subtracting the brightest sources. The zoomed in window in the left side of the image also shows that the faint sources are not removed at all. However, clustered calibration could remove all the faint sources almost perfectly and has generated much less artefacts after subtracting for the brightest sources in the right side of the field of view. Moreover, the residual noise of the clustered calibration follows a symmetric zero mean Gaussian distributions with a standard deviation of 4.2 mJy, while the one from the un-clustered calibration has an asymmetric Gaussian distribution with mean and standard deviation equal to -1.2 and 5.3 mJy, respectively. Taking to account that the simulated noise distribution is a zero mean Gaussian distribution with a standard deviation of 3 mJy, the better performance of the clustered calibration compared to the un-clustered one is evident. }\label{Figure111}
\end{figure*}

The residual images of the clustered calibration with $Q=27$ number of source clusters, and the un-clustered calibration for $K=52$ individual sources are shown by Fig. \ref{Figure111}. In the right side of the residual image of the un-clustered calibration there exist artificial stripes caused by over and under estimating the brightest sources of the field of view. That shows the problematic performance of the un-clustered calibration. However, clustered calibration has generated much less artificial effects after subtracting these sources. On top of that, the zoomed in window in the left side of the images of  Fig. \ref{Figure111}  show that the faint sources are not removed by the un-clustered calibration at all, while being almost perfectly subtracted by the clustered calibration. Moreover, the residual noise of the clustered calibration follows a symmetric zero mean Gaussian distributions with a standard deviation of 4.2 mJy, while the one from the un-clustered calibration has an asymmetric Gaussian distribution with mean and standard deviation equal to -1.2 and 5.3 mJy, respectively. Taking to account that the simulated noise distribution is a zero mean Gaussian distribution with a variance of 3 mJy, the superior performance of the clustered calibration compared to the un-clustered one is evident.

As the final conclusion of this simulation, calibrating below the noise level, clustered calibration always performs better than the un-clustered calibration. This is regardless of the sky model and is only based on the fact that solving for individual sources with very poor signals is impractical. Nevertheless, when some sources are very close to each other, the sky corruption on their signals would be exactly the same and there is no point in solving for every of them individually.

\section{Conclusions}\label{summary}
In this paper, we demonstrate the superior performance of ``clustered calibration'' compared to un-clustered calibration especially in calibrating sources that are below the calibration noise level. The superiority is in the sense of having more accurate results by the enhancement of SNR as well as by the improvement of computational efficiency by reducing the number of directions along which calibration has to be performed. 

In a ``clustered calibration'' procedure, sky sources are grouped into some clusters and every cluster is calibrated as a single source. That replaces the coherencies of individual sources by the total coherency of the cluster. Clustered calibration is applied to these new coherencies that carry a higher level of information compared with the individual ones. Thus, for the calibration of sources below the noise level it has a considerably better performance compared to un-clustered calibration. Another way of looking at clustering is to consider the distribution of source flux densities, i.e. the number of sources vs. the source flux density curve. Regardless of the intrinsic flux density distribution, clustering makes  the number of clusters vs. the cluster flux density curve more uniform, thus yielding superior performance.  An analytical proof of this superiority, for an arbitrary  sky model, is presented using MCRLB and SINR analysis.

KLD and LRT are utilized to detect the optimum number of clusters, for which the clustered calibration accomplishes its best performance. A model for estimating SINR of clustered calibration is also presented by which we could find the optimum number of clusters at low computational cost. 

Divisive hierarchical as well as Weighted K-means clustering methods are used to exploit the spatial proximity of the sources. Simulation studies reveal clustered calibration's improved performance at a low SNR, utilizing these clustering algorithms. Both the clustering methods are hard clustering techniques which divide data to distinct clusters. However, we expect more accurate results using fuzzy (soft) clustering, which constructs overlapping clusters with uncertain boundaries. Application and performance of this type of clustering for clustered calibration will be explored in future work.

\section*{acknowledgment}
The first author would like to gratefully acknowledge NWO grant 436040 and to thank N. Monshizadeh for the useful discussions. We also thank the referee for a careful review and valuable comments that helped enhance this paper.

\bibliographystyle{mn2e}
\bibliography{references}
\bsp
\label{lastpage}
\end{document}